\documentclass{aa}  

\makeatletter

\let\AA@old@journalname\aa@journalname
\let\AA@old@manuscriptname\aa@manuscriptname
\let\AA@old@AALogo\AALogo
\let\AA@old@today\today
\newlength\AA@old@fboxrule
\newlength\AA@old@fboxsep

\newcommand*\AA@disableTitleBanner{%
	\renewcommand*\aa@journalname{}
	\renewcommand*\aa@manuscriptname{}
	\renewcommand*\AALogo{}
	\def\today{}
	\setlength{\AA@old@fboxrule}{\fboxrule}
	\setlength{\AA@old@fboxsep}{\fboxsep}%
	\setlength{\fboxrule}{0pt}%
	\setlength{\fboxsep}{0pt}%
}

\newcommand*\AA@restoreTitleBanner{%
	\let\aa@journalname\AA@old@journalname
	\let\aa@manuscriptname\AA@old@manuscriptname
	\let\AALogo\AA@old@AALogo
	\let\today\AA@old@today
	\setlength{\fboxrule}{\AA@old@fboxrule}%
	\setlength{\fboxsep}{\AA@old@fboxsep}%
}

\let\AA@old@maketitle\maketitle
\renewcommand*\maketitle{%
	\AA@disableTitleBanner
	\AA@old@maketitle
	\AA@restoreTitleBanner
}

\makeatother

\usepackage{fancyhdr}
\fancypagestyle{plain}{%
	\fancyhf{}                 
	\fancyfoot[C]{\thepage}    
	
}
\usepackage[utf8]{inputenc}
\usepackage[varg]{txfonts}  
\usepackage{bm}
\usepackage{amssymb}
\usepackage{amsmath}
\usepackage{mathrsfs}
\usepackage{graphicx}
\usepackage{tabularx}
\usepackage{enumitem}
\usepackage{color}
\usepackage{subcaption}
\usepackage{hyperref}
\usepackage{booktabs}
\usepackage{multirow}
\usepackage{array}
\usepackage{upgreek}
\usepackage{appendix}
\makeatletter

\usepackage{natbib}
\usepackage{aliascnt}

\defcitealias{2025A&A...702A.208R}{Paper I}

\hypersetup{
	colorlinks = true,
	linkcolor = blue,
	citecolor = blue,
	filecolor = blue,
	urlcolor = blue
}

\def\apjl{ApJL}
\def\apjs{Astrophys. J. Supp.}

\def\aap{Astron. Astrophys.}

\newcolumntype{K}[1]{>{\centering\arraybackslash}p{#1}}

\let\OLDthebibliography\thebibliography
\renewcommand\thebibliography[1]{
	\OLDthebibliography{#1}
	\setlength{\parskip}{0pt}
	\setlength{\itemsep}{0pt plus 0.3ex}
}

\allowdisplaybreaks

\titlerunning{Black hole spin-up in proto-stellar clusters}

\title{Spin-up and spin distribution of stellar black holes grown by gas accretion in proto-stellar clusters}

\author{Zacharias Roupas\inst{1,2}}

\institute{Dipartimento di Fisica ``G. Occhialini'', 
	Universit\'a degli Studi di Milano-Bicocca, Piazza della Scienza 3, 20126 Milano, Italy
	\email{zacharias.roupas@unimib.it}
	\and
	Istituto Nazionale di Fisica Nucleare (INFN), Sezione di Milano-Bicocca, 
	Piazza della Scienza 3, 20126 Milano, Italy
}

\date{}

\abstract{
	Proto-stellar clusters, likely progenitors of globular clusters, are extremely compact with typical masses of $\sim 10^6\,{\rm M}_\odot$ and sizes of $\sim 1\,{\rm pc}$, as has recently been revealed by James Webb Space Telescope  observations at $z\sim 10$. Sufficiently high compactness can provide a time window for early-formed stellar black holes (BHs) to accrete primordial gas. We developed a semi-analytic model  to follow BH spin-up and determine the final spin distribution of stellar BHs that grow in mass via gas accretion within compact gaseous proto-stellar clusters. The velocity shear within a BH's sphere of influence induces the formation of an accretion disk that is repeatedly disrupted by stochastic perturbations to the BH motion. We assumed low initial BH spins of $a_{*,{\rm ini}} = 0.01$, consistent with stellar-evolution models with efficient angular-momentum transport, and we restricted initial BH masses to values below the upper BH mass gap, $m_{\rm BH,ini} < 55\,{\rm M}_\odot$. Our analysis shows a strong BH spin-mass correlation, obtained within $\sim 10 \,{\rm Myr}$ when gas is depleted.
	Low-spin BHs, $a_{*} \leq 0.3$, are predominantly low-mass,  $m_{\rm BH} \lesssim 25\,{\rm M}_\odot$, in contrast to high-spin BHs, $a_{*} \geq 0.7$, which are predominantly high-mass, $m_{\rm BH} \gtrsim 65\,{\rm M}_\odot$. Notably, there exist also low-spin, high-mass outliers with $\sim 1$ mass-gap BH per cluster expected to have $a_{*} \sim 0.1$. The general trend, however, expressed by the median spin as a function of final BH mass, is well fit by a high-spin saturating exponential with a transition mass of $\sim 50\,{\rm M}_{\odot}$. For $m_{\rm BH} \geq 100\,{\rm M}_\odot$ the median spin is $\bar{a}_{*} \sim 0.90$, with the central $68\%$ of the distribution spanning $a_{*} \sim 0.70 - 0.96$, in striking agreement with the estimated spins of the BH components of the gravitational-wave signal GW231123. These spin values persist up to the highest masses generated by our mechanism, $m_{\rm BH} \sim 10^3\,{\rm M}_\odot$.
}

\begin{document}
	
	\maketitle
	
	\section{Introduction}\label{sec:intro}
	
	In \cite{2025A&A...702A.208R} (hereafter \citetalias{2025A&A...702A.208R}) we proposed a channel for the black hole (BH) mass growth via gas accretion in proto-stellar clusters to values within the upper BH mass gap, induced by the physics of a pair-instability supernova (PISN) \citep{2016A&A...594A..97B,2017MNRAS.470.4739S,2021ApJ...912L..31W}, $60\,{\rm M}_\odot \lesssim m_{\rm BH} \lesssim 130\,{\rm M}_\odot$, and up to the intermediate-massive-black-hole (IMBH) regime of $m_{\rm BH} \sim 10^3\,{\rm M}_\odot$. 
	Gas-rich, compact, low-metallicity, massive stellar clusters provide ideal conditions for the mass growth of stellar BHs via accretion. Mass segregation, which drives the massive stellar progenitors of BHs toward the dense center of the cluster \citep{Spitzer_1987degc}, is accelerated in the presence of ambient gas remaining from the first stellar formation event, at timescales of $\lesssim 1\,{\rm Myr}$ \citep{2014MNRAS.441..919L}. Moreover, in a lower-metallicity environment stellar winds get weaker, allowing for longer gas-depletion timescales in the cluster, despite PISN feedback, for sufficiently compact clusters \citepalias{2025A&A...702A.208R}. 
	
	Recent observations by the James Webb Space Telescope (JWST) revealed five proto-stellar cluster candidates in the Cosmic Gems arc galaxy at high redshift, $z = 10.2$, \citep{2024Natur.632..513A}, which present low metallicity and are remarkably compact. Their typical masses are $\sim 10^6\,{\rm M}_\odot$ and half-light radii $\sim 1\, {\rm pc}$. 
	Such proto-stellar clusters are plausibly the progenitors of globular clusters (GCs) \citep{2015MNRAS.454.1658K,2017MNRAS.469L..63R,2018ApJ...869..119E}. These JWST observations provide observationally motivated cluster parameters for our models.
	
	The semi-analytic framework of stellar-BH growth in such environments, introduced in \citetalias{2025A&A...702A.208R}, incorporates gas depletion by stellar feedback, distinct BH formation times depending on stellar progenitors and formation channel, dynamical friction, stochastic stellar encounters, turbulence, and gas accretion.
	Here we extend this model to include the BH spin evolution. This calculation is essential for identifying our proposed channel regarding the astrophysical origin of gravitational-wave (GW) signals, among different stellar-BH mass growth channels. These include the repeated BH merger channel \citep{2002MNRAS.330..232C,2006ApJ...637..937O,2017PhRvD..95l4046G}, and gas accretion in active galactic nucleus (AGN) disks \citep{2007MNRAS.374..515L,2012MNRAS.425..460M,2018ApJ...859L..25Y,2020ApJ...901L..34Y} as well as in the cores of proto-galactic haloes \citep{2020ApJ...903L..21S,2025arXiv250808558B}.
	
	Identifying the origin of heavy BHs $\sim 10^2-10^3\,{\rm M}_\odot$ is a timely, active area of research  
	\citep{2017PhRvD..95l4046G,2019A&A...632L...8R,2020PhRvD.102d3002B,2020ApJ...904L..13R,Kimball_2020,2020ApJ...903L..21S,2021MNRAS.505..339M,2022MNRAS.512..884R,2023MNRAS.526..429A,2024A&A...688A.148T,2024Sci...384.1488F,2024AJ....167..191P,2025PhRvD.111f3039B,2025ApJ...988...15L,2025arXiv250808558B,2025ApJ...994L..37K,2025ApJ...993L..54G,2025ApJ...993L..30L}.
	The GW experiments, LIGO-Virgo-KAGRA Collaborations, have revealed a significant population of BHs within the upper BH mass gap \citep{2023PhRvX..13d1039A}. The most recent such GW signal, GW231123 \citep{2025ApJ...993L..25A}, involves components that are heavy, $103_{-52}^{+20}\, {\rm M}_\odot$, $137_{-17}^{+22}\, {\rm M}_\odot$, and that have high spins, $0.9_{-0.19}^{+0.10}$, $0.8_{-0.51}^{+0.20}$. These values are difficult to explain within the repeated-merger scenario, which predicts a narrow peak at $\sim 0.7$ \citep{2017ApJ...840L..24F}. We show here that these mass and spin values naturally emerge via accretion in gas-rich proto-stellar clusters.
	
	The BH motion is repeatedly deflected by stochastic gravitational perturbations -- two-body encounters and turbulent gas motions -- that reorient the BH trajectory.
	Angular-momentum inflow is induced by the velocity shear within the BH's sphere of influence, which can produce an accretion disk. These successive episodes lead to an accretion regime, where a broad range of spin values, $\sim 0-0.97$,  is possible, with a strong spin-mass correlation, as we demonstrate below.
	
	In the next section we review the cluster and BH motion model of \citetalias{2025A&A...702A.208R}.
	In Sect. \ref{sec:spin_disk} we describe the BH spin evolution and accretion disk models. Section \ref{sec:results} presents our results, and we conclude in Sect. \ref{sec:conclusion}. 
	
	\section{Cluster physical model}\label{sec:model}
	
	We adopted the same semi-analytic model as we introduced in  \citetalias{2025A&A...702A.208R} for the proto-stellar cluster and the BH motion. We briefly summarize the main elements in this section and refer the reader to \citetalias{2025A&A...702A.208R} for further details.
	
	\subsection{Gas depletion timescale}
	
	In gaseous proto-stellar clusters, stellar winds and supernova (SN) explosions inject energy into the ambient gas, driving gas depletion. Theoretical arguments and observational evidence indicate that the depletion timescale depends primarily on the compactness of the cluster \citep{2016A&A...587A..53K,2025A&A...702A.208R},
	\begin{equation}
		C \equiv \frac{M/10^5{\rm M}_\odot}{r_{\rm h}} ,
	\end{equation}
	rather than on the cluster mass alone. A simple theoretical understanding follows from noting that the binding energy scales as $U \propto M^2/R \sim C \cdot M$, and therefore increases linearly with the total mass for a fixed initial compactness, while the total stellar feedback energy scales as $E_{\rm feed} \propto M$ regardless of compactness. Consequently, compactness -- not mass -- primarily determines the relative energy balance between feedback and gravitational binding.      
	The star formation efficiency is also crucial in setting the depletion timescale, as it regulates both the available gas reservoir and the total energy released by stellar feedback. A third key parameter is metallicity, which influences stellar wind luminosities and stellar lifetimes, and thereby the onset of SN feedback. In this work we focus on low-metallicity clusters ($Z \sim 0.01 Z_{\odot}$) motivated by high-redshift JWST observations of the Cosmic Gems arc galaxy. 
	
\begin{figure}[tbp]
		\centering
		\includegraphics[width=0.9\columnwidth]{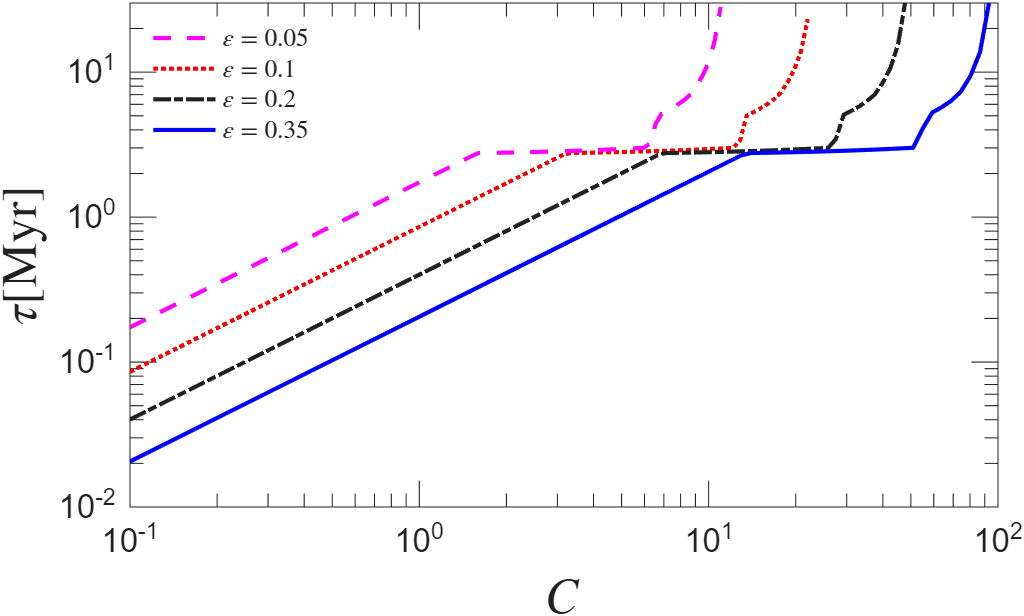}
		\caption{Depletion timescale, $\tau$, with respect to the compactness, $C$, of a gaseous stellar cluster of any total mass at low metallicity ($\sim 0.01Z_\odot$) for different possible star formation efficiencies, $\varepsilon$.}
		\label{fig:tau_dep}
	\end{figure}

	In Fig.~\ref{fig:tau_dep}, we display  the depletion timescale, $\tau$, as a function of compactness for several values of $\varepsilon$. This calculation is based on our detailed model presented in \citetalias{2025A&A...702A.208R}, where corresponding figures for solar and sub-solar metallicities are also provided. In our model $\tau$ is defined as the time when the cumulative energy from stellar winds and SN equals the binding energy. We took into account the stellar lifetimes dependence on stellar masses, extrapolating the tabulated data, referring to low-metallicity stars, of \citet{2019A&A...627A..24G}. Stellar-wind and SN energy injection was modeled based on a careful consideration of observations and numerical data \citep{2001A&A...369..574V,2002ApJ...577..389K,2008A&ARv..16..209P,2013A&A...550A..49K,2017A&A...603A.118R,2020MNRAS.499..873S}. 
	We observe in Fig.~\ref{fig:tau_dep} a characteristic stall at 
	\begin{equation}\label{eq:tau_stall}
		\tau_{\rm stall} \approx 2.9{\rm Myr} 
	\end{equation} 
	that marks the onset of PISNe for the adopted parameters. The range of compactness values for the observed proto-stellar clusters in the Cosmic Gems arc galaxy -- the same range as is used in our simulations -- corresponds to this stall timescale.
	
	\subsection{Black hole motion}
	
	In order to simulate the motion of BHs, we implemented the same model as in \citetalias{2025A&A...702A.208R}, briefly summarized as follows. 
	We assumed background gas and stellar Plummer density profiles, taking into account the expansion of the cluster due to the mass loss, the higher extension of the gas component relative to the stellar one, the gas temperature profile, and the gas cooling over time. We adopted an exponential law of gas depletion \citep{2007MNRAS.380.1589B}, 
	\begin{equation}
		\dot{M}_{\rm gas} = -\frac{M_{\rm gas}}{\tau},
	\end{equation} 
	since the stellar feedback processes, responsible for gas depletion, generate energy proportional to the cluster mass.
	We terminated each run once $99\%$ of the initial gas mass had been depleted. In practice, the BH mass and spin evolution stalled even earlier, when $\sim 95\%$ of the gas had been removed.
	
	At each simulation run, we generated a BH population from a Salpeter BH initial mass function \citep{2019ApJ...878L...1P} sampling positions and velocities from the Brownian probability density of \citet{2021A&A...646A..20R}. Each BH was created at a different time depending on the zero-age main sequence (ZAMS) progenitor mass. 
	We classified the BHs into several types depending on the ZAMS progenitor mass assumed to drive the related formation channel: a core-collapse-SN (CCSN), direct collapse of low or high ZAMS mass, and pulsational-pair-instability-SN (PPISN) of low or high ZAMS mass. We applied a recoil kick at birth to the BH remnants of CCSN that injected $\sim 70\%$ of such BHs out of the cluster.
	
	\begin{figure}[tbp]
		\centering
		\includegraphics[width=0.95\columnwidth]{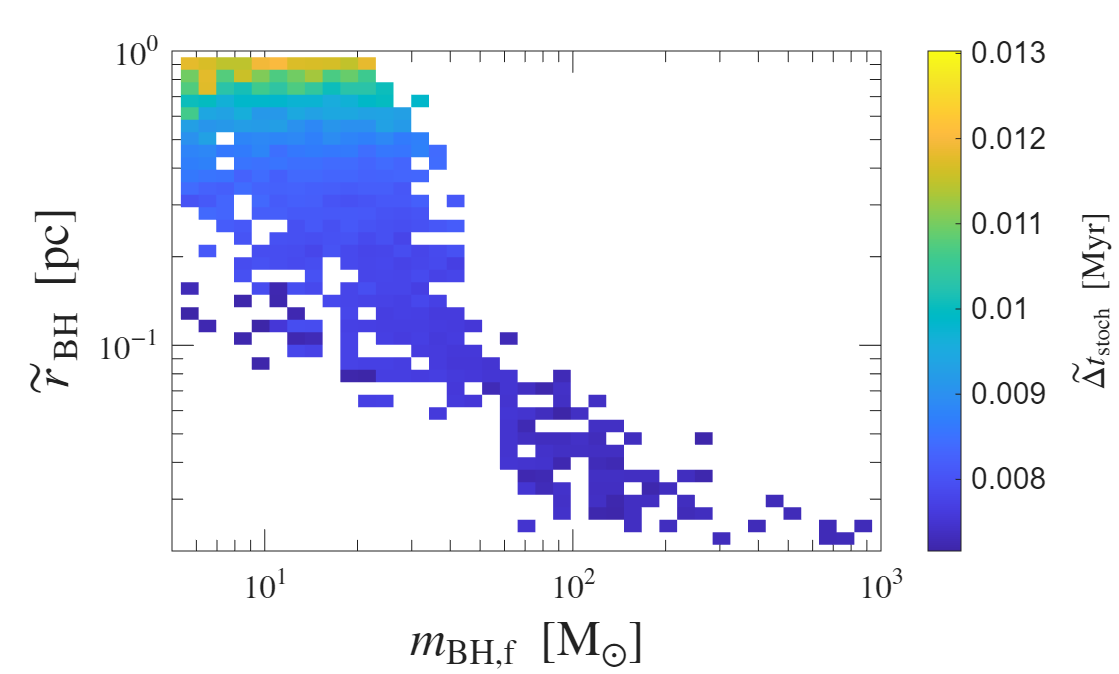}
		\caption{Contour of the median stochastic timestep, $\tilde{\Delta t}_{\rm stoch}$, with respect to the median BH position, $\tilde{r}_{\rm BH}$, and the final BH mass, $m_{\rm BH,f}$, for an indicative run of our typical cluster, $M_{\star} = 10^6 \,{\rm M}_\odot$, $r_{c,\star} = 1 \,{\rm pc}$, $\varepsilon = 0.35$.}
		\label{fig:Dt_stoch_r_m_contour2D}
	\end{figure}
	
	We evolved all BHs in the evolving background gas and stellar profiles, integrating the BHs equations of motion with a standard drift-kick scheme. The deterministic forces -- gravity and dynamical friction from the background -- were implemented using an adaptive Runge-Kutta method, while stochastic velocity increments due to both stellar perturbations and gas turbulencewe were applied at adaptive timesteps, $\Delta t_{\rm stoch}$, as is described in detail in \citetalias{2025A&A...702A.208R}. 
	The local gravitational dynamical timescale, $\tau_{\rm grav}(r_{\bullet}) = \sqrt{r_{\bullet}^3/(G M_{\rm enc}(r_{\bullet}))} \sim 1/\sqrt{G\left\langle \rho_{\rm tot}(r_{\bullet})\right\rangle}$, is the typical orbital timescale of the local perturber population. It is therefore a natural choice for setting the reference scale of the timestep, $\Delta t_{\rm stoch}$, for applying stochastic kicks. We further imposed consistency bounds, which are discussed in detail in Appendix \ref{app:dt_stoch}. In Fig. \ref{fig:Dt_stoch_r_m_contour2D} we plot the contour of the median $\tilde{\Delta t}_{\rm stoch}$ with respect to the median BH position and the final BH mass for an indicative run of our typical cluster. It is $\tilde{\Delta t}_{\rm stoch} = \mathscr{O} (10^{-2}\,{\rm Myr})$ and can become as low as $\sim 0.007\,{\rm Myr}$ for more heavy BHs, $m_{\bullet,{\rm f}} \gtrsim 50\,{\rm M}_\odot$, which move on average deeper in the cluster core. 
	
	We adopted an isotropic hot-type accretion, whose typical representative is Bondi-Hoyle accretion \citep{Merritt_book} given the dense, hot, turbulent gaseous environment.
	The spherically symmetric accretion rate is \citep{2002apa..book.....F}
	\begin{equation}\label{eq:dm}
		\dot{m}_{\rm feed} = \pi \rho_\text{gas} v_\text{rel} R_\text{acc}^2
		,\quad
		v_\text{rel} = \sqrt{v_\bullet^2 + c_s^2}.
	\end{equation} 
	The appropriate value of the accretion radius, $R_{\rm acc}$, depends on the relative amount of gas and the radial position of the accretor \citep{2001MNRAS.324..573B}.
	If the gas dominates the cluster potential at a certain BH position, $r_{\bullet}$, the accretion rate is determined by a tidal-lobe accretion radius
	\citep{1971ARA&A...9..183P},
	\begin{equation}
		R_\text{tid}(r_{\bullet}) \sim 0.5 \left( \frac{m_{\rm BH}}{M(r<r_{\bullet})} \right)^{1/3} r_{\bullet},
	\end{equation}
	where $M(r<r_{\bullet})$ is the total mass of the cluster within the BH radial position. 
	Otherwise, when gas is less dominant, the appropriate accretion radius is the Bondi-Hoyle radius, $R_{\rm B}$:
	\begin{equation}
		R_\text{B} = \frac{2 G m_{\rm BH}}{v_\text{rel}^2}.
	\end{equation}
	We assumed the accretion radius to be the smaller of the two at any instant in time \citep{2001MNRAS.324..573B},
	\begin{equation}\label{eq:R_acc}
		R_{\rm acc}(t) = \min \left\lbrace R_{\rm B}(t), \, R_{\rm tid}(t) \right\rbrace
		,\end{equation}
	adopting the most conservative assumption. 
	Furthermore, we adopted a maximum threshold for the accretion rate of
	\begin{equation}\label{eq:dm_acc}
		\dot{m}_{\rm acc}(t) = \min\left\lbrace \dot{m}_{\rm feed}(t), 10 \dot{m}_{{\rm Edd},0} \right\rbrace
		,\end{equation}
	where $
	\dot{m}_{{\rm Edd},0} = 4\pi G m_{\rm BH} m_p/\eta_0 \sigma_T c 
	$
	is the reference Eddington rate value for the radiative efficiency, $\eta_0 = 0.057$. 
	This is based on the results of 
	\cite{2016MNRAS.459.3738I}, who showed that for $1 < \dot{m}_{\rm feed} / \dot{m}_{\rm Edd,0} < 100$ the mean accretion rate cannot exceed $\left\langle \dot{m}_{\rm acc} \right\rangle \lesssim 10 \dot{m}_{\rm Edd,0}$.
	We adopted this constant cap as a simple, conservative assumption, since \cite{2016MNRAS.459.3738I} reported that there occur episodic bursts of accretion rate comparable to $\dot{m}_{\rm feed}$ that violate the general cap for the mean rate. In practice, in the physical conditions we consider here and for $m_{\rm BH} \lesssim 10^3{\rm M}_{\odot}$ we have $0 < \dot{m}_{\rm feed}(t)/\dot{m}_{\rm Edd,0} < 10$, automatically satisfying the cap.
	
	\section{BH-spin evolution and accretion disk}\label{sec:spin_disk}
	
	Our primary goal is to quantify the possible spin-up of stellar BHs through accretion in proto-stellar clusters. 
	A stellar BH is formed with low spin from its progenitor star if there is efficient angular-momentum transfer.
	The magnetic Tayler instability may efficiently couple the stellar core to its envelope \citep{2002A&A...381..923S}, enabling angular momentum to be transported outward during core contraction and thereby preventing the core from spinning up. Simulations indicate that the BH formed at collapse is expected to have a very low natal spin, with a typical value of $a_{*} \simeq 0.01$ \citep{2019ApJ...881L...1F}, where the dimensionless spin is $a_{*} \equiv J_{\rm BH} c/(G m_{\rm BH}^2)$ and $J_{\rm BH}$ the BH spin.
	We adopted this value for the initial BH spin magnitudes. 
	
	Regarding the initial BH spin orientations, we assumed an isotropic distribution inherited from an isotropic distribution of stellar spin axes. This assumption is consistent with observational analyses \citep{2010MNRAS.402.1380J,2023ApJ...944...39H}. It is also intuitively justified, at least for nonrotating clusters, by the turbulent nature of star-forming environments. We note, nevertheless,  that rotating clusters with sufficient rotational energy may induce partially stellar spin alignment \citep{2017NatAs...1E..64C}. We do not consider this case here.
	
	\subsection{Disk formation}\label{sec:disk}
	
	The BH motion is repeatedly perturbed at stochastic timesteps, as is discussed in Appendix \ref{app:dt_stoch}.
	The acquired BH azimuthal velocity induces apparent transverse velocity gradients within the accretion sphere ($|\bm{r} - \bm{r}_{\bullet}| \leq R_{\rm acc}$) as the sphere gets advected along the BH trajectory.
	Numerical simulations of angular-momentum accretion, driven by velocity and density gradients, show the formation of an accretion disk  \citep{1995A&A...295..108R}.
	More recently, \citet{2025ApJ...979...61T} demonstrated that an off-axis (``lateral'') Bondi-Hoyle-Lyttleton flow, characterized by an antisymmetric transverse velocity field, naturally generates angular momentum and produces a persistent disk.
	
	In our system the transverse velocity gradient is generated within the accretion radius by the BH azimuthal velocity, gained after a perturbation event, with $|\Delta v_{\perp}| \approx 2 |\omega_{\rm BH}| R_{\rm acc}$ (perpendicular to the position vector $\bm{r}_{\bullet}$) on the induced orbital plane, where 
	\begin{equation}
		\bm{\omega}_{\rm BH} = \frac{1}{r_\bullet^2}\,\bm{r}_{\bullet} \times \bm{v}_{\bullet} .
	\end{equation}
	The velocity shear across $R_{\rm acc}$ produces a net angular momentum as the apparent gas transverse velocity in the BH rest frame differs between the inner (proximate to the cluster center)  and outer hemispheres, creating a shear flow opposite to the BH's orbital motion. This asymmetric inflow provides angular-momentum injection within the BH's accretion radius $R_{\rm acc}$. At the BH orbital plane, this shear specific  angular momentum of the gas is $\bm{j}_0 \approx - 2 \bm{\omega}_{\rm BH} R_{\rm acc}^2$. Volume-averaging over all planes within $R_{\rm acc}$ (using $R_{\rm plane} = R_{\rm acc} \cos\theta$, $\theta$ measured from the orbital plane) and weighting by the volume element $\propto \cos^2\theta$, we obtain the geometric prefactor: $2\int_{-\pi/2}^{\pi/2}\cos^4\theta d\theta / \int_{-\pi/2}^{\pi/2}\cos^2\theta d\theta = 3/2$. Thus, we get the total specific shear angular momentum injected into the accretion radius,
	\begin{equation}\label{eq:j}
		\bm{j} \approx -\frac{3}{2} \bm{\omega}_{\rm BH} R_{\rm acc}^2,
	\end{equation}
	where the minus sign indicates that the shear generates retrograde inflow relative to BH orbital motion. The accretion radius is defined in Eq. (\ref{eq:R_acc}).
	Both the orientation and approximate magnitude of our estimate, Eq. (\ref{eq:j}), is consistent with numerical simulations of angular-momentum accretion driven by velocity gradients  \citep{1995A&A...295..108R,2025ApJ...979...61T}.
	
	Note that all BH-dependent quantities are local and time-dependent, such as $\bm{j} = \bm{j}(\bm{r}_{\bullet}, t)$, although we omit this dependence to simplify our notation.
	There is also an angular momentum component ($\bm{j}_{\rm grad}$) induced by the density gradient within $R_{\rm acc}$. For the smooth cluster density profiles in our 
	models, this is negligible because $\bm{j}_{\rm grad} \approx (\Delta m_{\rm in-out}/m_{\rm acc}) \bm{\omega}_{\rm BH} R_{\rm acc}^2$ with $\Delta m_{\rm in-out}/m_{\rm acc} \ll 1$ (the mass difference between the inner and outer accretion hemispheres over the total sphere mass). It is straightforward to verify that for our typical parameters of a Plummer sphere, a BH inside the Plummer radius and a disk of any radius, $R_{\rm d} < 10^{-3}\,{\rm pc}$, we get $j_{\rm grad}/j \lesssim 10^{-3}$. We therefore neglected the $j_{\rm grad}$ contribution to improve computational efficiency.

	In our system, a rotationally supported disk can only form if the gas circularizes inside the BH's radius of influence but outside the innermost stable circular orbit (ISCO),
	\begin{equation}\label{eq:R_condition}
		R_{\rm ISCO} < R_{\rm circ} \leq R_{\rm acc} .
	\end{equation}
	We defined the circularization radius as
	\begin{equation}
		R_{\rm circ} = \frac{j^2}{Gm_{\rm BH}},
	\end{equation}
	which, using eq (\ref{eq:j}) and $\Omega_{{\rm K},\bullet} \equiv \sqrt{Gm_{\rm BH}/r_{\bullet}^3}$, can be written as
	\begin{equation}\label{eq:R_circ}
		\frac{R_{\rm circ}}{R_{\rm acc}} =      \frac{9}{4} \left(\frac{R_{\rm acc}}{r_{\bullet}} \right)^3
		\left(\frac{\omega_{\rm BH}}{\Omega_{{\rm K},\bullet}} \right)^2 .
	\end{equation}

	\subsection{Spin magnitude evolution}
	
	In our system, the accretion rate,
	\begin{equation}
		\dot{m} \equiv \frac{\dot{m}_{\rm acc}}{\dot{m}_{\rm Edd,0}},
	\end{equation}
	depends sensitively not only on the BH mass, but also on the BH position and velocity. Therefore, it varies for the same BH during the evolution of the system. We have
	\begin{equation}
		0 \leq \dot{m}(t) \leq 10
	\end{equation}
	for each BH.
	A geometrically thin-disk approximation is valid for stellar BHs if $\dot{m} \lesssim 0.3$ \citep{2021ARA&A..59..117R}. For higher accretion rates, the disk is mildly geometrically thick or ``slim'' up to our upper accretion cap, $\dot{m} \leq 10$ \citep{2009ApJS..183..171S}. Thus, we assumed that the disk of each BH is 
	\begin{align*}
		& \mbox{thin, if } \dot{m}(t) < 0.3, \\
		& \mbox{slim, if } 0.3 \leq \dot{m}(t)  \leq 10, 
	\end{align*}
	and used the appropriate model at each instant of time.

	A slim disk has a lower radiative efficiency, $\eta$, than a thin disk \citep{2009ApJS..183..171S}. We adopted the simple prescription \citep{2016MNRAS.459.3738I}
	\begin{equation}
		\eta_{\rm eff} = \eta_{\rm thin} \frac{1}{1 + \frac{\dot{m}}{10/3}}.
	\end{equation}
	This gives $\eta_{\rm eff} \approx \eta_{\rm thin}$ at $\dot{m} \lesssim 0.3$. At the cap $\dot{m} = 10$, we get $\eta_{\rm eff} = 0.014 - 0.08$ ($\eta_{\rm thin} = 0.057-0.32$) for spin zero and one, respectively. 
	
	We calculated the BH mass increase rate as
	\begin{equation}
		\dot{m}_{\rm BH} = (1 - \eta_{\rm eff}(a_{*})) \dot{m}_{\rm acc} 
	\end{equation}
	and the BH-spin magnitude evolution as
	\begin{equation}\label{eq:dadt}
		\frac{da_{*}}{dt} = \frac{\dot{m}_{\rm acc}}{m_{\rm BH}}\left( \Lambda - 2a_{*} (1 - \eta_{\rm eff})  \right) ,
	\end{equation}
	with
	\begin{equation}
		\Lambda \equiv 
		\left\lbrace
		\begin{array}{ll}
			\Lambda_{\rm thin}, & \dot{m} < 0.3 \\
			\Lambda_{\rm slim}, & \dot{m} \geq 0.3
		\end{array}
		\right.
		.\end{equation}
	The values of $\Lambda_{\rm thin}\equiv \Lambda(R_{\rm ISCO})$ and 
	$\eta_{\rm thin} \equiv 1 - \mathcal{E}(R_{\rm ISCO})$ 
	are given by the standard thin-disk expressions \citep{1972ApJ...178..347B,1974ApJ...191..507T}. 
	
	We introduced a phenomenological relation to estimate the effective $\Lambda_{\rm slim}$ for slim disks,
	\begin{equation}
		\Lambda_{\rm slim} = f_\Lambda(H/R) \Lambda_{\rm thin},\;
		f_{\Lambda}^{\rm min} \leq f_\Lambda (H/R) \leq 1,
	\end{equation}
	where $f_\Lambda(H/R)$ decreases linearly. This linear dependence is motivated by general relativistic magnetohydrodynamic (GRMHD) simulations \citep{2010MNRAS.408..752P}, which  show a linear decrease in the specific angular momentum accreted by the BH with increasing disk thickness, $H/R$, of a few percent. We estimated $H/R$, modeling the simulation data of the slim-disk models of \cite{2011A&A...527A..17S} with a simple saturating prescription,
	\begin{equation}
		H/R = 0.5\frac{ (\dot{m}/10)^{1/2} }{1 + (\dot{m}/10)^{1/2} },
		\quad 0.3\leq \dot{m} \leq 10.
	\end{equation}
	This gives reasonable values of $H/R = \{0.07,\, 0.12,\, 0.25\}$ for $\dot{m} = \{0.3,\, 1,\, 10\}$, respectively. 
	We have $f_{\Lambda}^{\rm max} = 1$ for the thin disk, $\dot{m} = 0.3$. Instead, the parameter $f_{\Lambda}^{\rm min}$ is defining the minimum possible angular momentum inflow corresponding to our maximum allowed accretion rate, $\dot{m} = 10$. The direct correspondence between $f_{\Lambda}$ and $a_{*,{\rm max}}$, which can be derived by the equilibrium condition of (\ref{eq:dadt}), allows for the estimation of $f_{\Lambda}^{\rm min}$ by estimating $a_{*,\rm{max}}$ for slim disks, $\dot{m} =10$. The analytical calculations of \citet{2011A&A...532A..41S} suggest that for slim disks at super-Eddington accretion rates it is $a_{*,\rm{max}}\approx 0.99$ close to the Thorne limit. Instead, GRMHD simulations of
	slim disks show that we may get a significantly lower equilibrium spin, $a_{*,\rm{max}}\approx 0.93$ \citep{2004ApJ...602..312G}. This corresponds to $f_{\Lambda}^{\rm min} = 0.95$. We adopted this value as a conservative assumption on the spin-up process and performed sensitivity checks. We find that our results are robust and practically unaffected for values even down to $f_{\Lambda}^{\rm min} = 0.88$, corresponding to the extreme case of an equilibrium  spin, $a_{*,\rm{max}}=0.90$, at $\dot{m}=10$ for every single accretion cycle at this rate. Even for $f_{\Lambda}^{\rm min} = 0.80$ our derived spin-mass correlation, including the high-spin plateau, remains robust, while the saturation spin drops only by $\lesssim 4\%$.
	
	We note that if a magnetically arrested disk (MAD) \citep{1974Ap&SS..28...45B} forms, the accreting BH spins down due to angular momentum extraction via the Blandford-Znajek mechanism  \citep{2011MNRAS.418L..79T}.   
	\citet{2023ApJ...954L..22R} have estimated that for thick disks the equilibrium spin drops down to $a_{*,{\rm max}} \approx 0.8,\, 0.2$ for $\dot{m} = 1,\, 10$, respectively. 
	However, in the dense stochastic environment of a gaseous proto-cluster the maintenance of a long-lived coherent MAD is uncertain. In addition, \citet{2023ApJ...954L..22R} suggest that the spin equilibration timescale for MAD accretion attains the values $t_{\rm eq}^{\rm  MAD}\approx 45,\,10,\,4.5\,{\rm Myr}$ for $\dot{m} = 1, \,5, \,10$, respectively. 
	Our mass-gap BHs have been accreting during their evolution at a rate of $\dot{m} \sim 1-3$ or less, while only the IMBH with $m_{\rm BH} \sim 10^3\,{\rm M}_\odot$ may have reached up to $\dot{m} \sim 10$. In our model the cluster gas is depleted within $13\,{\rm Myr}$; therefore, MAD accretion is certainly irrelevant for our mass-gap BHs ($< 150\,{\rm M}_\odot$) for which $t_{\rm eq}^{\rm  MAD} > t_{\rm depletion}$. Most importantly, as we discuss in Appendix \ref{app:dt_stoch}, the BHs are disrupted at a timescale of $\Delta t_{\rm stoch} \approx 10^{-2}\,{\rm Myr}$. 
	Therefore, both for mass-gap and IMBHs, even if a MAD disk forms, it cannot remain coherent after a few disruption events and does not have sufficient time during its lifetime to cause any significant spin-down.  
	Overall, our model assumes that MAD accretion is negligible. 
	
	\subsection{Spin-disk orientation}
	
	The accretion disk, if formed, gets erratically disrupted, repeatedly, every timestep, $\Delta t_{\rm stoch}$ (dynamically calculated as the system evolves), when a stochastic kick due to stellar encounters and gas turbulence is applied. Since the disk orientation depends on the BH motion, each disruption event drives a disk reorientation. 
	
	In order to evaluate the spin-disk relative orientation after every disruption event, we need the disk angular momentum, $J_{\rm d} \equiv J(R_{\rm d})$, where
	\begin{equation}
		J(R) \equiv \int_{R_{\rm ISCO}}^{R} 2\pi r \Sigma(r) \sqrt{G m_{\rm BH} r} dr.
	\end{equation}
	Therefore, we need a model for the surface density, $\Sigma$, and the disk radius, $R_{\rm d}$, which does not necessarily coincide with the circularization radius.
	We calculated the disk surface density, adopting the Shakura-Sunjaev model \citep{1973A&A....24..337S,2002apa..book.....F}, 
	\begin{equation}\label{eq:Sigma}
		\Sigma(R) = 4\cdot 10^{-13} \,\frac{{\rm M}_{\odot}}{r_{\rm g}^2}
		\left(\frac{\alpha}{0.1} \right)^{-\frac{4}{5}}
		\left(\frac{m_{\rm BH}}{50\,{\rm M}_\odot} \right)^{\frac{11}{5}}
		\left(\frac{R}{r_{\rm g}} \right)^{-\frac{3}{4}}
		\dot{m}^{\frac{7}{10}}
		f(R),
	\end{equation}
	where
	$
	f(R) = (1 - ( R/R_{\rm ISCO} )^{-1/2} )^{1/4}
	$
	and the gravitational radius is $r_{\rm g} = Gm_{\rm BH}/c^2$.
	For thin disks we adopted the viscosity parameter $\alpha_{\rm thin} = 0.1$.
	For slim disks, the Shakura-Sunjaev surface density can be a good approximation using an effective viscosity parameter of $\alpha_{\rm eff}= 0.01$ \citep{2011A&A...527A..17S}.
	Indeed, we verified that for $\alpha = 0.01$, the Shakura-Sunyaev surface density matches the slim-disk GRRMHD simulations of \cite{2011A&A...527A..17S} to within a factor of order unity.  Therefore we adopted this as a sufficient analytic approximation. We also performed exploratory sensitivity checks over the range $\alpha_{\rm eff} = 0.001 - 0.1$. The general trend of our derived spin-mass correlation is robust. A notable result is that higher $\alpha_{\rm eff}$ produces a somewhat broader BH spin distribution for the more massive BHs. 
	Especially for $\alpha_{\rm eff} = 0.1$ the IMBHs ($\sim 10^3{\rm M}_\odot$) exhibit reduced lower and upper spin boundaries and a nearly uniform spin distribution within this range: $a_{*} \sim 0.3-0.8$. For $m_{\rm BH} \lesssim 800\, {\rm M}_{\odot}$, however, there is no reduction in spin boundaries, but only a slight increase in their spread.
	Nevertheless, we emphasize that  $\alpha_{\rm eff} = 0.1$ is not strictly self-consistent within our slim-disk treatment, because it lowers the effective optical depth, thereby shifting the validity of the model below our super-Eddington regime for slim disks \citep{2011A&A...527A..17S}. 
	
	Assuming $R_{\rm d} \gg R_{\rm ISCO}$ we have
	\begin{equation}\label{eq:J_ratio}
		\frac{J_{\rm d}}{2J_{\rm BH}} = 
		9\cdot 10^{-7} a_{*}^{-1} 
		\dot{m}^{\frac{7}{10}}
		\left(\tfrac{\alpha}{0.01} \right)^{-\frac{4}{5}}
		\left(\tfrac{m_{\rm BH}}{50\,{\rm M}_\odot} \right)^{\frac{6}{5}}
		\left(\tfrac{R_{\rm d}}{10^4 r_{\rm g}} \right)^{\frac{7}{4}}.
	\end{equation}
	In order to define the disk radius, $R_{\rm d}$, we need first to introduce three characteristic radii. 
	Firstly, we defined the outer radius, $R_{\rm out}$, as the maximum outer radius at which torques have enough time to be communicated, before the BH motion gets disrupted by a stochastic kick. It is defined by the condition
	$
	t_{\rm w}(R_{\rm out}) = \Delta t_{\rm stoch}
	$,
	which gives
	\begin{equation}\label{eq:R_out}
		\frac{R_{\rm out}}{r_{\rm g}} =
		\left\lbrace
		\begin{array}{ll}
			1.5\cdot 10^8 \left( \frac{\Delta t_{\rm stoch}}{10^{-3}\,{\rm Myr}}\right)^{\frac{4}{5}}        
			\left( \frac{m_{\rm BH}}{50 {\rm M}_{\odot}}\right)^{-\frac{24}{25}} &,\;\mbox{thin} \\
			3.5\cdot 10^8 \left( \frac{\Delta t_{\rm stoch}}{10^{-3}\,{\rm Myr}}\right)^{\frac{2}{3}}
			\left( \frac{H/R}{0.1}\right)^{\frac{2}{3}}     
			\left( \frac{m_{\rm BH}}{50 {\rm M}_{\odot}}\right)^{-\frac{2}{3}} &,\;\mbox{slim}
		\end{array}
		\right. 
		.\end{equation}
	We define $t_{\rm w}$ and discuss timescales in Appendix \ref{app:disk}. It should be $R_{\rm d} \leq R_{\rm out}$; otherwise, there is not enough time for the torque to be communicated out to $R_{\rm d}$ within the stochastic timestep, $\Delta t_{\rm stoch}$.
	
	The Toomre radius is defined by the condition $Q(R_{\rm Q}) = 1$, where the Toomre parameter is $Q(R) = c_{\rm s,disk} \Omega_{\rm K,disk}(R)/(\pi G \Sigma(R))$. Stable disks are constrained by the condition $Q(R) > 1$; that is, $R < R_{\rm Q}$. We have
	\begin{equation}
		\frac{R_{\rm Q}}{r_{\rm g}} =
		\left\lbrace
		\begin{array}{ll}
			1.7\cdot 10^{10}
			\left( \frac{m_{\rm BH}}{50 {\rm M}_{\odot}}\right)^{-\frac{52}{45}} 
			\dot{m}^{-\frac{22}{45}}
			&,\;\mbox{thin} \\
			2.9\cdot 10^{9}
			\left( \frac{H/R}{0.1}\right)^{\frac{4}{5}} 
			\left( \frac{m_{\rm BH}}{50 {\rm M}_{\odot}}\right)^{-\frac{24}{25}} 
			\dot{m}^{-\frac{14}{25}} &,\;\mbox{slim}
		\end{array}
		\right. 
		.\end{equation}
	It should also be $R_{\rm d} \leq R_{\rm Q}$; otherwise, the disk is unstable.
	
	At last, we defined the ``breaking'' radius $R_{\rm break}$. Numerical evidence suggests that the disk (either thin or slim) breaks into rings above a threshold radius \citep{2012ApJ...757L..24N, 2015MNRAS.448.1526N},
	\begin{equation}
		\frac{R_{\rm break}}{r_{\rm g}} = 
		\left( \frac{4 a_*}{3\alpha (H/R)}\left| \sin\theta_{\rm rel} \right|\right)^{2/3}\,
		,\quad
		45^o\lesssim \theta_{\rm rel } \lesssim 135^o
		,\end{equation}   
	for such a severe misalignment \citep{2012ApJ...757L..24N}. 
	The angle $\theta_{\rm rel}$ is the relative angle between the initial disk angular momentum (at the time a stochastic kick is applied) and BH-spin orientation,
	\begin{equation}
		\cos \theta_{\rm rel} = - \hat{\bm{\omega}}_{\rm BH}\cdot \hat{\bm{J}}_{BH} |_{t=t_{\rm ini}}
		,\end{equation}
	since $\hat{\bm{J}}_{\rm disk} = - \hat{\bm{\omega}}_{\rm BH}$.
	It should be $R_{\rm d} \leq R_{\rm break}$ when $45^o\lesssim \theta_{\rm rel } \lesssim 135^o$; otherwise, the disk breaks into rings.
	
	Thus, $R_{\rm d}$ should be smaller than all four characteristic radii, $R_{\rm circ}$, $R_{\rm out}$, $R_{\rm Q}$, and $R_{\rm break}$; that is, we defined
	\begin{equation}\label{eq:R_d}
		R_{\rm d} = \min\{R_{\rm circ}, R_{\rm out}, R_{\rm Q}, R_{\rm break}\}
		,\end{equation}
	provided always that $R_{\rm ISCO} < R_{\rm circ} \leq R_{\rm acc}$, so that a disk is able to form. The $R_{\rm break}$ is considered only when $45^o\lesssim \theta_{\rm rel } \lesssim 135^o$; otherwise, it is ignored.

	For $R_{\rm acc} = R_{\rm B}$, which is the typical case for BHs close the cluster center, the circularization radius is
	\begin{equation}\label{eq:R_circ_an}
		\frac{R_{\rm circ}}{r_{\rm g}} = 1.2\cdot 10^4
		\left( \tfrac{T_{\rm cl}}{10^4{\rm K}}\right)^{-4}
		\left( \tfrac{m_{\rm BH}}{50\,{\rm M}_\odot}\right)^3
		\left( \tfrac{r_{\bullet}}{0.1\,{\rm pc}}\right)^{-3}
		\left( \tfrac{\omega_{\rm BH}}{\Omega_{{\rm K},\bullet}}\right)^{2}
		\left(1 + \mathcal{M}_{\bullet}^2 \right)^{-4}
		,\end{equation}
	where the Mach number is $\mathcal{M}_{\bullet} = v_\bullet / c_{{\rm s,cl}}$, the cluster gas temperature is $4\cdot 10^3{\rm K} \lesssim T_{\rm cl}(t) \lesssim 1.2\cdot 10^4{\rm K}$, and we always assume ionized gas, $m_{\rm eff} = 0.6m_{\rm p}$, with $\gamma = 5/3$. Therefore for smaller BHs, $m_{\rm BH} < 10^3\,{\rm M}_\odot$, the smaller radius is most probably $R_{\rm circ}$, depending, however, on temperature and BH kinematics (with the exception of strong misalignment when $R_{\rm break}$ is smaller). For larger BHs the $R_{\rm out}$ may be the smallest.
	
	The spin-disk relative orientation depends on $J_{\rm d}/(2 J_{\rm BH})$, $\theta_{\rm rel}$, and the BH-alignment timescale, $t_{{\rm al},\bullet}$, with respect to the stochastic-kick disruption timestep, $\Delta t_{\rm stoch}$ (see Appendix \ref{app:disk}). 
	As our simulated system evolved, we recomputed $J_{\rm d}(t)$ and $J_{\rm BH}(t)$ dynamically each time a stochastic kick was applied and then we adopted the following prescription:
	\begin{itemize}[noitemsep,nolistsep]
		\item If $J_{\rm d}/(2 J_{\rm BH}) < 1$, the disk aligns with the BH spin if the relative inclination satisfies $\theta_{\rm rel} < \tfrac{\pi}{2}$ and counteraligns otherwise \citep{2005MNRAS.363...49K}. 
		\item If $J_{\rm d}/(2J_{\rm BH}) > 1$, the BH spin aligns with the disk for all initial orientations if $t_{{\rm al},\bullet} < \Delta t_{\rm stoch}$. If instead $t_{{\rm al},\bullet} > \Delta t_{\rm stoch}$ we assume steady misalignment \citep{2008MNRAS.385.1621K} and also substitute in eq. (\ref{eq:dadt}), $\Lambda = \cos(\theta_{\rm rel})\Lambda_{\rm thin}$ or $\cos(\theta_{\rm rel})\Lambda_{\rm slim}$, depending on the disk type.
	\end{itemize}
	Once the relative orientation was specified, we evolved the spin magnitude (\ref{eq:dadt}) for the time interval, $\Delta t_{\rm stoch}$, until a new disruption event occurred.
	
	We stress that slim disks, specifically the super-Eddington ones responsible for the high-spin/high-mass BHs, have a significantly shorter BH alignment timescale than thin disks, because the maximum torque is exerted at a shorter distance from the BH. We calculated this quantitatively in detail in Appendix \ref{app:disk}, with the final expression given in Eq.~(\ref{eq:t_al_slim}). In practice, the regime $J_{\rm d}/(2J_{\rm BH}) > 1$, which allows for a net spin-up, may occur when a BH has grown to $m_{\rm BH} > 50\,{\rm M}_\odot$, accretes at $\dot{m}> 1$ (slim disk), and reaches a disk radius of $\mathscr{O}(10^8 r_{\rm g})$. This follows from Eqs. (\ref{eq:J_ratio}), (\ref{eq:R_d}), and (\ref{eq:R_circ_an}), together with the fact that more massive BHs sink deeper into the center, $r_\bullet < 0.1\,{\rm pc}$. 
	For such BHs we typically get $t_{{\rm al},\bullet}^{\rm slim} \sim 10^{-2} \Delta t_{\rm stoch}$ (see Eq.~\ref{eq:t_al_slim}), and certainly $t_{{\rm al},\bullet}^{\rm slim} \lesssim 0.1 \Delta t_{\rm stoch}$ for BHs within the upper BH mass gap range. Therefore, most BHs at and above the mass gap threshold are expected to reach the maximum possible spin allowed by the present accretion geometry, while the rarer cases of $t_{{\rm al},\bullet} > \Delta t_{\rm stoch}$ (and significant $J_{\rm d}$) are already treated through our steady-misalignment prescription. 
	This distinguishes our process from the chaotic accretion for SMBHs \citep{2006MNRAS.373L..90K,2008MNRAS.385.1621K}: the underlying disk-formation and evolution physics are different for stellar BHs in clusters and for AGN disks. 
	We note, nevertheless, that strong spin-up (above $a_{*} > 0.3$) concerns a minority of the BH population, $\lesssim 10\%$. The majority of the population, corresponding mainly to lower BH masses, stays in the chaotic-accretion regime, and could spin down (those kinematically capable of forming disks) if the initial spin was high. Even so, the transition to high spins at high BH masses and the high-spin/high-mass plateau persists even for higher assumed initial BH spins.
	
	\begin{figure}[tbp]
		\centering
		\includegraphics[width=0.9\columnwidth]{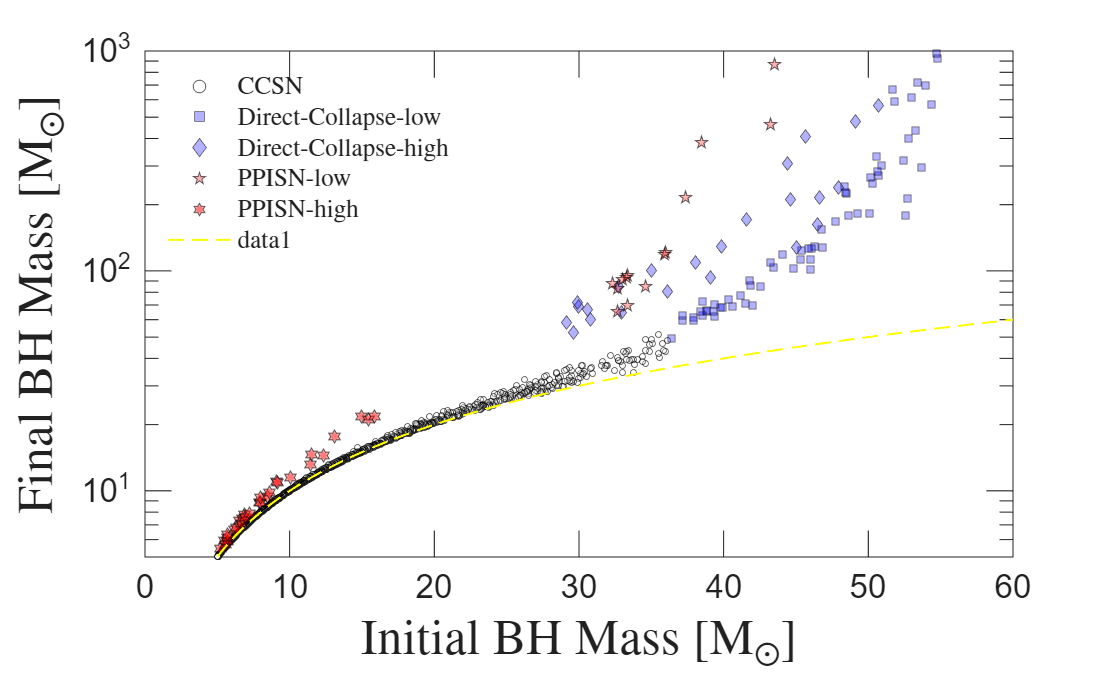}
		\caption{Mass scatter diagram for $M_{\star} = 10^6\,{\rm M}_\odot$, $r_{c,\star} = 1 \,{\rm pc}$, star formation efficiency $\varepsilon = 0.35$, and one run.}
		\label{fig:m_fin-ini}
	\end{figure}
	
	\begin{figure}[tbp]
		\centering
		\includegraphics[width=0.9\columnwidth]{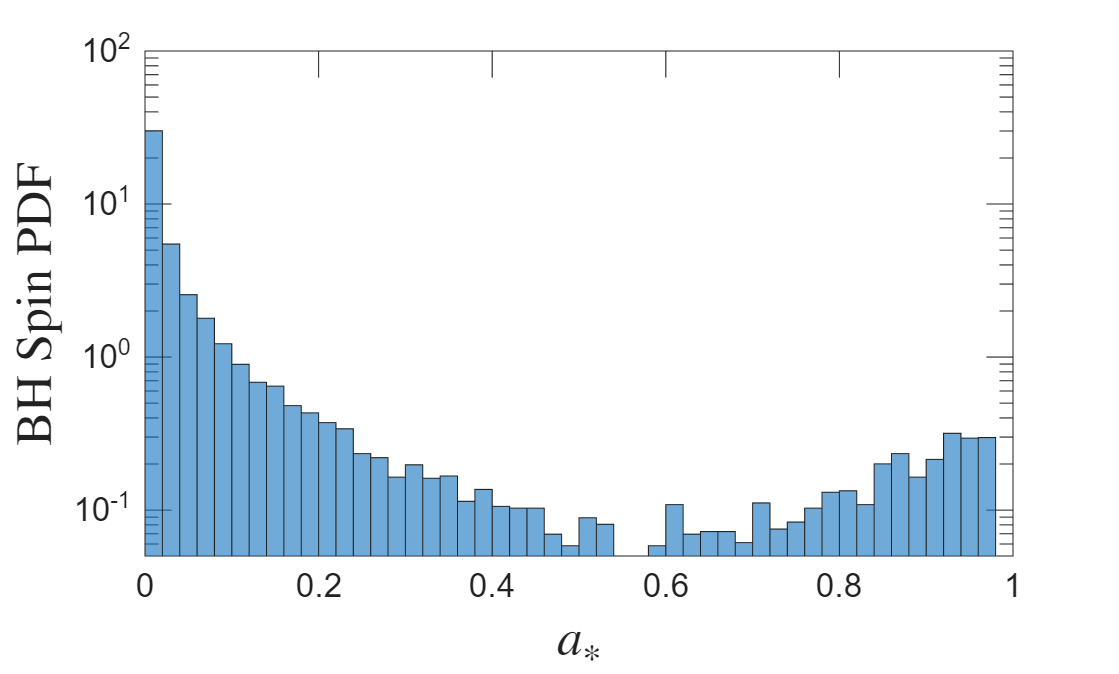}
		\caption{Final BH spin probability density function for $M_{\star} = 10^6\,{\rm M}_\odot$, $r_{c,\star} = 1 \,{\rm pc}$, star formation efficiency $\varepsilon = 0.35$, and ten runs.} 
		\label{fig:a_PDF}
	\end{figure}
	
	\section{Results}\label{sec:results}
	
	We calibrated our primary cluster parameters to be consistent with observations of proto-stellar clusters by JWST in the Cosmic Gems arc galaxy \citep{2024Natur.632..513A}. Their observed masses are in the range of $(1-4)\cdot 10^6\,{\rm M}_\odot$ and the half-light radii are $(0.7 - 1.7)\, {\rm pc}$. 
	Their observationally estimated age of $\sim 10-50\,{\rm Myr}$ suggests that they are possibly nearly at the final gas expulsion stage, with little or no gas remaining. However, the exact amount of gas remaining is uncertain, while the galaxy itself is gas-rich. We assumed a low metallicity of $0.01 Z_\odot$, in accordance with the low metallicity expected for the Cosmic Gems arc galaxy \citep{2024Natur.632..513A}.
	
	We adopted an initial total mass for the initial gas-rich cluster of $M = 3\cdot 10^6\,{\rm M}_\odot$ and a star formation efficiency of $\varepsilon = 0.35$, which combined give $M_{\star} = 10^6\,{\rm M}\odot$ for the stellar component, at the low-mass end of the observations. Such a high $\varepsilon$ represents the upper end, $\sim 0.3$, of observed star formation efficiencies in embedded clusters \citep{2003ARA&A..41...57L,2007MNRAS.380.1589B}, appropriate for the extreme densities and low metallicities in such massive, compact proto-stellar clusters as the Cosmic Gems ones.
	We cover the range of our theoretical Plummer radii of the stellar component at the end of gas depletion equivalent to $r_{c,\star} = 1.0-1.3 \,{\rm pc}$, which is the appropriate variable to compare with the observed half-light radii. 
	These Plummer radii originate in a more compact iniital gaseous configuration with initial half-mass radii (accounting for both gas and stellar components) of $r_{\rm h}^{(\rm ini)} \sim  0.6 - 0.8 \,{\rm pc}$ because of the cluster expansion during gas depletion. This range is ideal, since it corresponds to the stall in the depletion timescale (Fig. \ref{fig:tau_dep}), with similar $\tau \approx 2.9\,{\rm Myr}$ for the whole range of adopted initial compactness. Such a timescale gives a $99\%$ depletion by $13\,{\rm Myr}$ consistent with the observations that show little to no gas in the Cosmic Gems cluster with an age of $>10\,{\rm Myr}$. We terminated the simulations at this time of $99\%$ depletion. 
	
	\begin{figure*}[tb]
		\centering
		\begin{subfigure}{0.48\textwidth}
			\centering
			\includegraphics[width=1.0\textwidth]{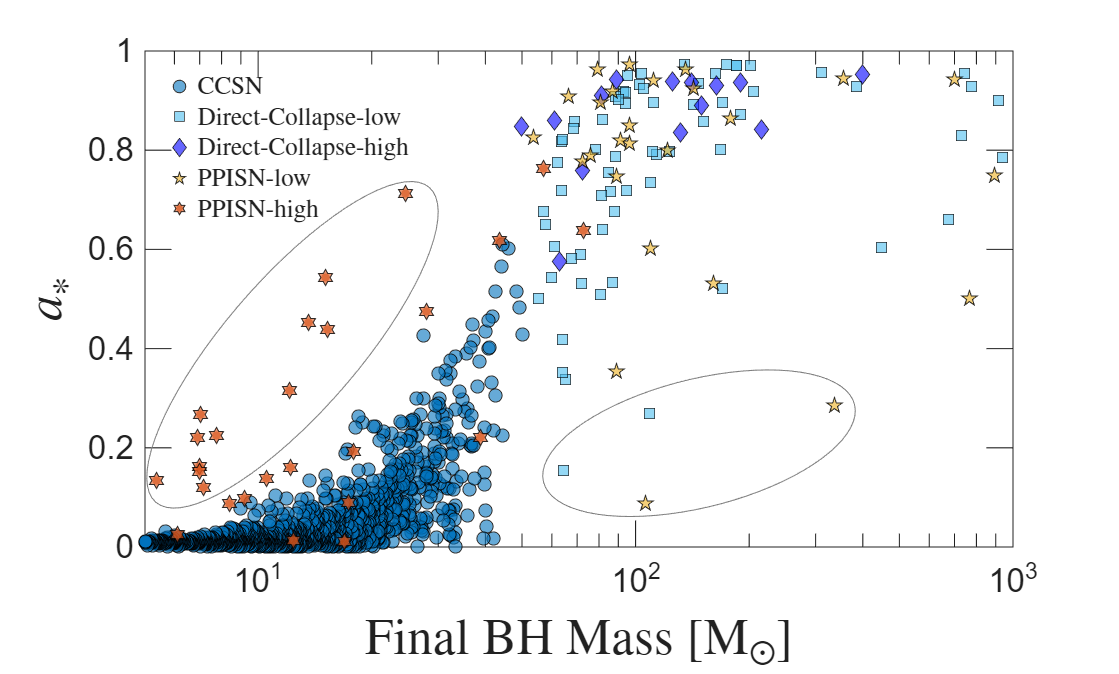}
			\caption{}
			\label{fig:a_m_fin_data}
		\end{subfigure}
		\begin{subfigure}{0.48\textwidth}
			\centering
			\includegraphics[width=1.0\textwidth]{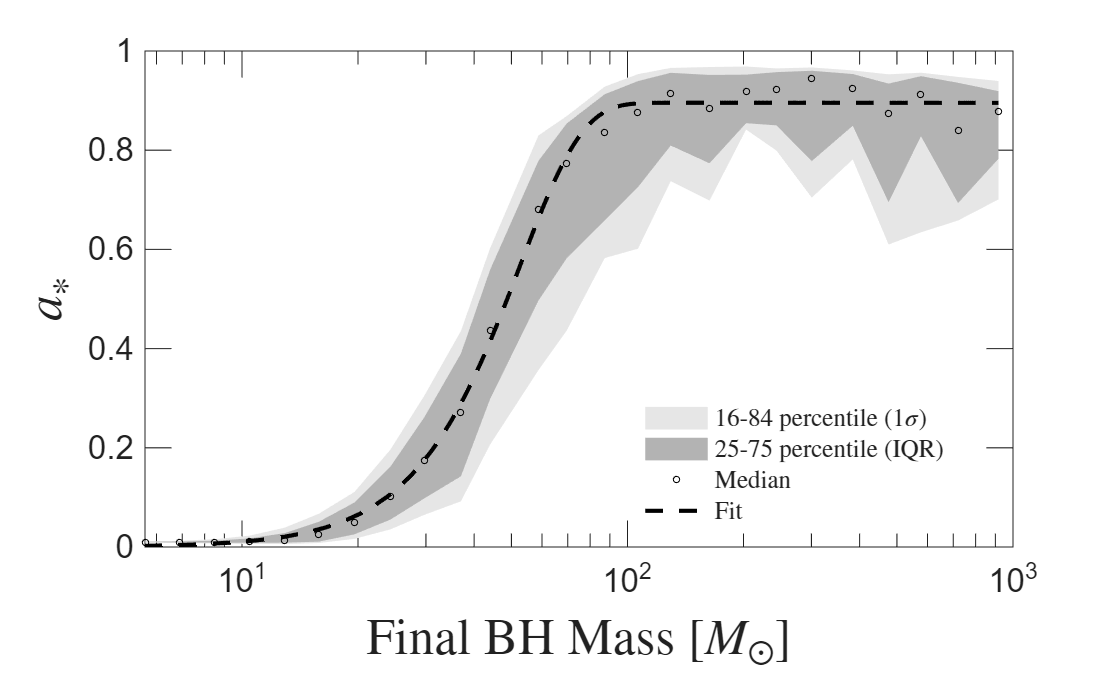}
			\caption{}
			\label{fig:a_m_fin}
		\end{subfigure}
		\caption{Final BH spin with respect to final BH mass for our typical cluster with $M_{\star} = 10^6\,{\rm M}_\odot$, $r_{c,\star} = 1 \,{\rm pc}$, and a star formation efficiency of $\varepsilon = 0.35$. (a) Scatter diagram for one indicative run. We have circled \{high-spin, low-mass\} and \{low-spin, high-mass\} outliers. (b) Percentiles, median, and fit relation, expressed by Eq.~(\ref{eq:a_med}), for ten runs.}
		\label{fig:a_m_both}
	\end{figure*}
	
	In Fig. \ref{fig:m_fin-ini} we depict the scatter diagram of final BH masses with respect to their initial values for one simulation run for visual clarity with $M_{\star} = 10^6\,{\rm M}_\odot$, $r_{c,\star} = 1\,{\rm pc}$. We signify the specific BH type based on its formation channel. The PPISN and direct-collapse BHs are those that can populate the upper BH mass gap and reach even IMBH masses, because they are formed earlier, originating in more massive stars than the CCSN BHs. We have further investigated in detail the BH mass growth in \citetalias{2025A&A...702A.208R}, specifying the cluster parameters for which the upper BH mass gap can be populated and an IMBH of mass $\sim 10^3\,{\rm M}_\odot$ can be generated. 
	
	A bimodality of spin distribution is evident in Fig. \ref{fig:a_PDF}, with $91\%$ of BHs attaining a low spin of $a_{*} \leq 0.3$ and $5\%$ a high spin of $a_{*} \geq 0.7$. 
	We calculated that among all of the BHs with a low spin, $90\%$ of them have a mass of $m_{\rm BH} < 22.7 \,{\rm M}_\odot$, and that among all of the BHs with a high spin, $90\%$ of them have a mass of $m_{\rm BH} > 63.6\,{\rm M}_\odot$, with the maximum median $\bar{a}_{*}$ achieved at $m_{\rm BH} \approx 100\,{\rm M}_\odot$.
	
	Figure \ref{fig:a_m_fin_data} shows the final BH spin corresponding to BH final masses for one run. The PPISN BHs of high ZAMS mass progenitors grow in spin more rapidly because they are born earlier than all BHs. They present a distinct trend at lower masses of $\lesssim 20\,{\rm M}_\odot$; however, at higher masses, PPISN and direct-collapse BHs present a similar trend with high spins of $\gtrsim 0.7$. The CCSN BHs do also spin up, but not above $a_{*} \lesssim 0.7$, although they do not grow significantly in mass. We stress that there exist outliers with low spin at high masses among the direct-collapse-high and PPISN-low BHs. On the contrary, the PPISN-high BHs tend to be outliers, with a high spin and a low mass. We discuss outliers in more detail in the following.
	
	\begin{table*}[tb]
		\caption{\label{tab:a_m_fit}
			Fit parameters of BH spin and saturation value for several clusters.}
		\centering
		\begin{tabular}{c || c c | c c || c c c c }                     
			$M_{\star}{[{\rm M}_\odot]}$
			&                               
			$r_{c,\star} {[{\rm pc}]}$
			&                       
			$C_{\rm ini}$
			&
			$\tau {[{\rm Myr}]}$
			&                       
			$m_{\rm BH}^{\rm max} {[{\rm M}_\odot]}$
			&
			$a_{*}^{\rm max}$
			&
			$\{a_{\rm M}, m_{\rm T}{[{\rm M}_\odot]}, \beta\}$
			&                       
			$\bar{a}_{*}^{\rm sat}{[1\sigma]}$
			\\[0.2ex]
			\toprule
			\multirow{2}{*}{{\color{blue} $10^6$}}
			&                               
			\color{blue} $1.0$
			&                               
			\color{blue} $44.0$
			&                       
			\color{blue} $2.86$
			&
			\color{blue} $984$
			&
			\color{blue} $0.973$
			&                       
			\color{blue} $\{0.90, 53, 2.7\}$
			&
			\color{blue} $0.90^{+0.06}_{-0.20}$
			\\[0.5ex]
			&                               
			\color{blue} $1.3$
			&                               
			\color{blue} $37.5$
			&                       
			\color{blue} $2.87$
			&
			\color{blue} $168$
			&                       
			\color{blue} $0.965$
			&                       
			\color{blue} $\{0.85,56,2.6\}$
			&
			\color{blue} $0.85^{+0.07}_{-0.30}$
			\\
			\midrule
			\multirow{2}{*}{$3.5\times 10^5$}
			&                               
			$0.6$
			&                               
			$25.0$
			&                       
			$2.83$
			&
			$1090$
			&                       
			$0.971$
			&                       
			$\{0.80,43,2.8\}$
			&
			$0.79^{+0.09}_{-0.14}$
			\\[0.5ex]
			&                               
			$0.8$
			&                               
			$18.5$
			&                       
			$2.81$
			&
			$175$
			&                       
			$0.967$
			&                       
			$\{0.77, 56, 2.7\}$
			&
			$0.78^{+0.12}_{-0.17}$
			\\
			\midrule
			\multirow{2}{*}{$10^5$}
			&                               
			$0.3$
			&                               
			$13.5$
			&                       
			$2.76$
			&
			$1041$
			&                       
			$0.902$
			&                       
			$\{0.70,33,2.7\}$
			&
			$0.65^{+0.10}_{-0.10}$
			\\[0.5ex]
			&                               
			$0.5$
			&                               
			$10.4$
			&                       
			$2.14$
			&
			$178$
			&                       
			$0.891$
			&                       
			$\{0.64, 50, 3.3\}$
			&
			$0.66^{+0.12}_{-0.13}$
		\end{tabular}
		\tablefoot{The first column lists the total mass of the stellar component, which originates in a gaseous cluster with total masses of $3\cdot 10^6$, $10^6$,  $3\cdot 10^5$ ${\rm M}_\odot$, respectively, given our adopted star formation efficiency, $\varepsilon = 0.35$. Subsequent columns list the stellar component Plummer radius, the initial compactness of the gaseous cluster before gas depletion, the depletion timescale, the maximum BH mass, the maximum BH spin, and the fit parameters of the median spin (\ref{eq:a_med}). The last column lists the median saturating spin with its $1\sigma$ percentile for $m_{\rm BH} \geq 100\,{\rm M}_{\odot}$. We chose compactness values that give similar $\tau\lesssim \tau_{\rm stall}$ (Eq.~\ref{eq:tau_stall}) for all cases, while covering the two primary cases of BH masses: either the maximum possible BH mass, $m_{\rm BH}^{\rm max} \approx 10^3 \, {\rm M}_{\odot}$, is generated, or the upper BH mass gap is marginally populated, $m_{\rm BH}^{\rm max} \approx 150 \, {\rm M}_{\odot}$. Blue entries indicate a stellar cluster mass and size consistent with JWST observations of the Cosmic Gems clusters.}
	\end{table*}
	
	\begin{table*}[tb]
		\caption{\label{tab:a_m_outlier}
			Number of high-mass/low-spin BH outliers per cluster for several clusters.}
		\centering
		\begin{tabular}{c || c || c c | c c | c c }                     
			\multicolumn{2}{c||}{}
			&
			\multicolumn{2}{c|}{$60\,{\rm M}_\odot \leq m_{\rm BH}^{\rm fin} < 150\,{\rm M}_\odot$}
			&
			\multicolumn{2}{c|}{$150\,{\rm M}_\odot \leq m_{\rm BH}^{\rm fin} \leq 500\,{\rm M}_\odot$}
			&
			\multicolumn{2}{c}{$500\,{\rm M}_\odot < m_{\rm BH}^{\rm fin} \leq 10^3\,{\rm M}_\odot$}
			\\[0.1ex]
			\midrule
			$M_{\star}{[{\rm M}_\odot]}$
			&                               
			$r_{c,\star} {[{\rm pc}]}$
			&                       
			{\small $\bar{n}(a_{*} \leq 0.1)$}
			&
			{\small $\bar{n}(0.1 < a_{*} \leq 0.3)$}
			&                       
			{\small $\bar{n}(0.1 \leq a_{*} < 0.3)$}
			&
			{\small $\bar{n}(0.3 \leq a_{*} \leq 0.5)$}
			&
			{\small $\bar{n}(0.1 \leq a_{*} < 0.3)$}
			&
			{\small $\bar{n}(0.3 \leq a_{*} \leq 0.5)$}
			\\[0.2ex]
			\toprule
			\multirow{2}{*}{{\color{blue} $10^6$}}
			&                               
			\color{blue} $1.0$
			&                               
			\color{blue} $0.70$
			&                       
			\color{blue} $2.80$
			&
			\color{blue} $0.40$
			&
			\color{blue} $1.20$
			&                       
			\color{blue} $0.10$
			&
			\color{blue} $0.50$
			\\[0.5ex]
			&                               
			\color{blue} $1.3$
			&                               
			\color{blue} $1.60$
			&                       
			\color{blue} $3.60$
			&
			\color{blue} $0.00$
			&
			\color{blue} $0.00$
			&                       
			\color{blue} $0.00$
			&
			\color{blue} $0.00$
			\\
			\midrule
			\multirow{2}{*}{$3.5\times 10^5$}
			&                               
			$0.6$
			&                               
			$0.00$
			&                       
			$0.20$
			&
			$0.05$
			&
			$0.30$
			&                       
			$0.00$
			&
			$0.09$
			\\[0.5ex]
			&                               
			$0.8$
			&                               
			$0.15$
			&                       
			$0.80$
			&
			$0.05$
			&
			$0.20$
			&                       
			$0.00$
			&
			$0.00$
		\end{tabular}
		\tablefoot{The cluster parameters correspond to the four first rows of Table \ref{tab:a_m_fit}. We denote as $\bar{n}$ the number of BHs per cluster within the corresponding mass bin and spin range.}
	\end{table*}

	The median $\bar{a}_{*}$ is well fit by a saturating exponential:
	\begin{equation}\label{eq:a_med}
		\bar{a}_{*}(m_{\rm BH}) = a_{\rm M} \left( 1 - e^{-\left(\frac{m_{\rm BH}}{m_{\rm T}}\right)^\beta}\right).
	\end{equation}
	The parameter $a_{\rm M}$ is the asymptotic median spin value at high masses, and the ``transition mass scale'', $m_{\rm T}$, designates the point of transition from low to high spin, with $\bar{a}_{*}(m_{\rm T})/a_{\rm M} = 0.63$. Remarkably, this function (\ref{eq:a_med}) is not specific to pre-assumed cluster mass and size values. It persists for a range of cluster masses and sizes, and not only for intense mass growth with $m_{\rm BH}^{\rm max} \approx 10^3\,{\rm M}_\odot$, but also even when the upper BH mass gap is only marginally populated, with $m_{\rm BH}^{\rm max}  \approx 150{\rm M}_\odot$. The fit parameter values for several clusters covering those two main cases of BH mass growth is shown in Table \ref{tab:a_m_fit}. 
	The cluster parameters affect the highest possible BH mass and only through this the maximum spin, preserving, however, the general trend of the spin-up equation, Eq. (\ref{eq:a_med}). The saturating median spin at high masses, $m_{\rm BH} \geq 100\,{\rm M}_\odot$, ranges from $\bar{a}_{*}^{\rm sat}\sim 0.65$ for a low cluster mass, $M_{\star} = 10^5\,{\rm M}_\odot$, to $\bar{a}_{*}^{\rm sat} \sim 0.90$ for a high cluster mass, $M_{\star} = 10^6\,{\rm M}_\odot$. The BH transition mass scale ranges slightly about $m_{\rm T}\sim 50\,{\rm M}_\odot$ with a median spin at transition, $\bar{a}_{*}(m_{\rm T})\sim 0.5$. The fit exponent is $\beta \sim 2.7$, with the exception of the least massive, least compact cluster case.
	
	In Fig. \ref{fig:a_m_fin} we illustrate the $1\sigma$ and IQR percentiles of the final BH spin with respect to the final mass, together with the median and its fit for the standard parameters $M_{\star} = 10^6\,{\rm M}_\odot$ and $r_{c,\star} = 1.0\,{\rm pc}$, typical representative values of Cosmic Gems clusters observed by JWST. For heavy BHs, $m_{\rm BH} \geq 100\,{\rm M}_\odot$, the $1\sigma$ range spans $a_{*} = 0.70 - 0.96$, with a saturating median spin of $\bar{a}_{*}^{\rm sat} = 0.90$. 
	
	These predicted spin values and corresponding masses encompass both components of GW231123 \citep{2025ApJ...993L..25A} with masses of $m_1 = 137_{-17}^{+22} \,{\rm M}_\odot$ and $m_2 = 103_{-52}^{+20} \,{\rm M}_\odot$ and spins of $a_{*,1} = 0.9_{-0.19}^{+0.10}$ and $a_{*,2} = 0.8_{-0.51}^{+0.20}$. The more massive component has a higher spin, consistent with our predicted mass-spin correlation, and both central spin values along with nearly their entire uncertainty ranges lie well within the predicted distribution for their respective masses. This agreement persists down to the less compact cluster case, $r_{c,\star} = 1.3 \,{\rm pc}$, as in Table \ref{tab:a_m_fit}. 
	
	The BHs grown by accretion in our cluster are expected to assemble into merging binary black holes (BBHs) through the standard dynamical channels of dense stellar systems; namely, exchange interactions involving existing binaries, three-body binary formation, and subsequent hardening through binary-single and binary-binary encounters (e.g., \citealt{2013LRR....16....4B,2016PhRvD..93h4029R}). Later cluster dynamics are expected to randomize spin orientations efficiently. The cosmic evolution of the proto-stellar cluster determines whether the spin-mass correlation of Eq.~(\ref{eq:a_med}) can survive as an observable imprint on BH spins. If the cluster evolves in the galactic halo into a GC, the imprint is likely to remain largely preserved, because tidal stripping and evaporation \citep{2003MNRAS.340..227B} reduce the cluster mass and density, thereby suppressing repeated mergers \citep{2019PhRvD.100d1301G}, which could modify the spin-mass relation. By contrast, if the cluster forms in, or later migrates to, the galactic nucleus and becomes part of a nuclear star cluster (NSC), the much higher escape speed can make hierarchical BH mergers efficient \citep{2019MNRAS.486.5008A}. The extent to which they wash out the original spin-mass correlation then depends on the cosmic epoch of cluster formation, the delay times for BBH assembly and successive mergers, and the growth history of the host NSC.

	We close this section with another result of our analysis. The stochastic nature of accretion allows for outliers with a low spin and a high mass. In Table \ref{tab:a_m_outlier} we list the number of BHs per cluster of outliers for several mass bins and spin ranges, for the massive clusters of the first four rows of Table \ref{tab:a_m_fit}. 
	For our typical representative of the Cosmic Gems clusters, there is an expected one mass-gap BH ($\sim 100\,{\rm M}_\odot$) per cluster with a low spin of $a_{*}\sim 0.1$. Massive BHs, $(150-10^3)\,{\rm M}_\odot$, with a spin lower than $a_{*} \lesssim 0.3$ are rarer. 
	Clusters with a lower mass ($M_{\star} \sim 10^5\,{\rm M}_\odot$) do not present significant outliers.
	In addition, as we have noted earlier, the PPISN-high BHs tend to be outliers -- sitting outside the $1\sigma$ distribution of Eq. (\ref{eq:a_med}) -- with higher spin despite their low mass. In particular those outliers with  a mass of $\leq 50{\rm M}_\odot$ are well fit by a function,
	$
	a_{*,{\rm low-mass}}^{\rm outliers} = A \log{\left( m_{\rm BH}^{\rm fin}\right)} - B,\quad m_{\rm BH}^{\rm fin} \leq 50\,{\rm M}_\odot
	$.
	The values of those fit parameters and the standard deviation of the spin, $\{A,B,\sigma_{a_{*}}\}$, are for the clusters of the first four rows of Table \ref{tab:a_m_fit}, respectively:
	$\{0.59,0.35,0.12\}$, $\{0.47,0.28,0.08\}$, $\{0.43,0.17,0.16\}$, and $\{0.39,0.20,0.11\}$. 
	
	\section{Conclusion}\label{sec:conclusion}
	
	We calculated the BH spin distribution generated by accretion in gaseous proto-stellar clusters within $\sim 10\,{\rm Myr}$ as gas is exponentially depleted by stellar feedback. We mainly considered parameter values consistent with JWST observations of five proto-stellar clusters in the Cosmic Gems arc galaxy \citep{2024Natur.632..513A} and also explored a broader parameter space. 
	
	Our accretion model incorporates the repeated disruption of BH accretion disks by gravitational perturbations in dense cluster environments. The disk orientation is determined by the acquired BH's azimuthal velocity, and therefore each disruption reorients the disk. Our major results are summarized in Tables \ref{tab:a_m_fit} and \ref{tab:a_m_outlier}, Fig. \ref{fig:a_m_fin}, and Eq.~(\ref{eq:a_med}). 
	
	This equation describes a BH spin distribution as a function of BH mass with a remarkably consistent functional form; it persists for all cluster parameter values investigated here, with representative ranges given in Table \ref{tab:a_m_fit}. This behavior, depicted graphically in Fig. \ref{fig:a_m_fin}, reveals a spin-mass correlation with three regimes: low spins ($a_{*} \lesssim 0.3$) for low BH masses, a steep increase with a transition mass scale of $\sim 50\,{\rm M}_\odot$ and a corresponding spin of $\sim 0.5$, and a saturating plateau for high masses. 
	
	For the typical Cosmic Gems cluster mass, $10^6\,{\rm M}_\odot$, and size, $1\,{\rm pc}$, the saturating spin reaches a median value of $\bar{a}_{*}^{\rm sat} = 0.90$ for $m_{\rm BH}\geq 100\,{\rm M}_\odot$, with the $1\sigma$ range covering $a_{*} = 0.70 - 0.96$ (see Table \ref{tab:a_m_fit}).
	Remarkably, these BH spin values and corresponding masses are in good agreement with GW231123 \citep{2025ApJ...993L..25A}, suggesting that some heavy BHs within the upper BH mass gap could have formed via accretion within the first compact massive stellar clusters, during their formation stage in the early Universe.
	
	Our predicted spin-mass correlation is most likely to survive in clusters that later evolve into GCs, where repeated mergers are expected to be inefficient. By contrast, if the cluster migrates to, or is formed inside, the galactic nucleus and becomes part of a NSC, repeated mergers may erase the correlation, depending mainly on the merger timescales and the growth history of the NSC.
	
	Finally, we stress that the stochastic nature of the process also underlies the presence of low-spin, high-mass outliers for sufficiently massive clusters representative of the Cosmic Gems population. Specifically, we expect $\sim 1$ mass-gap BH per cluster with a low spin of $a_{*} \lesssim 0.1$. Massive BHs ($m_{\rm BH} \gtrsim 500\,{\rm M}_\odot$) with a low spin of $a_{*} \lesssim 0.3$ are significantly rarer but cannot be excluded.
	
	Thus, our proposed BH spin and mass growth channel has two distinctive signatures: primarily, the spin-mass correlation of Fig. \ref{fig:a_m_fin}, mathematically expressed by the characteristic saturating exponential profile of the BH spin distribution, as in Eq.~(\ref{eq:a_med}); second, the presence of low-spin, high-mass BH outliers, which are difficult to explain with the repeated-merger channel. Possible important avenues for future work include (i) the analysis of GW data to identify potential candidates of our proposed channel based on these two signatures, and to distinguish them from those produced by the repeated-merger channel; (ii) the calculation of the contribution to the GW background generated by accreting BBHs in such gaseous proto-stellar clusters; and (iii) the calculation of IMBH binary merger rates in these environments, which is important for the planned LISA mission.

	\begin{acknowledgements}
		ZR is supported by the European Union's Horizon Europe Research and Innovation Programme under the Marie Sk\l{}odowska-Curie grant agreement No.~101149270--ProtoBH.
	\end{acknowledgements}
	
	\bibliography{aa58435-25}

@ARTICLE{1995A&A...295..108R,
	author = {{Ruffert}, M. and {Anzer}, U.},
	title = "{Bondi-Hoyle accretion simulations including velocity gradients.}",
	journal = {\aap},
	year = 1995,
	month = mar,
	volume = {295},
	pages = {108-112},
	adsurl = {https://ui.adsabs.harvard.edu/abs/1995A&A...295..108R}
}

@ARTICLE{2019ApJ...878L...1P,
	author = {{Perna}, Rosalba and {Wang}, Yi-Han and {Farr}, Will M. and {Leigh}, Nathan and {Cantiello}, Matteo},
	title = "{Constraining the Black Hole Initial Mass Function with LIGO/Virgo Observations}",
	journal = {\apjl},
	year = 2019,
	month = jun,
	volume = {878},
	number = {1},
	eid = {L1},
	pages = {L1},
	doi = {10.3847/2041-8213/ab2336},
	archivePrefix = {arXiv},
	eprint = {1901.03345},
	primaryClass = {astro-ph.HE},
	adsurl = {https://ui.adsabs.harvard.edu/abs/2019ApJ...878L...1P}
}

@ARTICLE{2021A&A...646A..20R,
	author = {{Roupas}, Zacharias},
	title = "{Gravitational Brownian motion as inhomogeneous diffusion: Black hole populations in globular clusters}",
	journal = {\aap},
	year = 2021,
	month = feb,
	volume = {646},
	eid = {A20},
	pages = {A20},
	doi = {10.1051/0004-6361/202039151},
	archivePrefix = {arXiv},
	eprint = {2006.12755},
	primaryClass = {astro-ph.GA},
	adsurl = {https://ui.adsabs.harvard.edu/abs/2021A&A...646A..20R}
}

@book{Merritt_book,
	Author = {David Merritt},
	Title = {Dynamics and Evolution of Galactic Nuclei (Princeton Series in Astrophysics)},
	Publisher = {Princeton University Press},
	Year = {2013},
	ISBN = {069112101X},
}

@ARTICLE{2025ApJ...979...61T,
	author = {{Tripathi}, Priyesh Kumar and {Chattopadhyay}, Indranil and {Joshi}, Raj Kishor},
	title = "{On Disk Formation around Isolated Black Holes via Stream Accretion}",
	journal = {\apj},
	year = 2025,
	month = jan,
	volume = {979},
	number = {1},
	eid = {61},
	pages = {61},
	doi = {10.3847/1538-4357/ad9b96},
	archivePrefix = {arXiv},
	eprint = {2412.04815},
	primaryClass = {astro-ph.HE},
	adsurl = {https://ui.adsabs.harvard.edu/abs/2025ApJ...979...61T}
}

@ARTICLE{2023ApJ...944...39H,
	author = {{Healy}, Brian F. and {McCullough}, P.~R. and {Schlaufman}, Kevin C. and {Kovacs}, Geza},
	title = "{A Study of Stellar Spins in 15 Open Clusters}",
	journal = {\apj},
	year = 2023,
	month = feb,
	volume = {944},
	number = {1},
	eid = {39},
	pages = {39},
	doi = {10.3847/1538-4357/acad7b},
	archivePrefix = {arXiv},
	eprint = {2301.10249},
	primaryClass = {astro-ph.SR},
	adsurl = {https://ui.adsabs.harvard.edu/abs/2023ApJ...944...39H}
}

@ARTICLE{2010MNRAS.402.1380J,
	author = {{Jackson}, R.~J. and {Jeffries}, R.~D.},
	title = "{Are the spin axes of stars randomly aligned within a cluster?}",
	journal = {\mnras},
	year = 2010,
	month = feb,
	volume = {402},
	number = {2},
	pages = {1380-1390},
	doi = {10.1111/j.1365-2966.2009.15983.x},
	archivePrefix = {arXiv},
	eprint = {0911.1075},
	primaryClass = {astro-ph.SR},
	adsurl = {https://ui.adsabs.harvard.edu/abs/2010MNRAS.402.1380J}
}

@ARTICLE{2017NatAs...1E..64C,
	author = {{Corsaro}, Enrico and {Lee}, Yueh-Ning and {Garc{\'\i}a}, Rafael A. and {Hennebelle}, Patrick and {Mathur}, Savita and {Beck}, Paul G. and {Mathis}, Stephane and {Stello}, Dennis and {Bouvier}, J{\'e}r{\^o}me},
	title = "{Spin alignment of stars in old open clusters}",
	journal = {Nature Astronomy},
	year = 2017,
	month = mar,
	volume = {1},
	eid = {0064},
	pages = {0064},
	doi = {10.1038/s41550-017-0064},
	archivePrefix = {arXiv},
	eprint = {1703.05588},
	primaryClass = {astro-ph.SR},
	adsurl = {https://ui.adsabs.harvard.edu/abs/2017NatAs...1E..64C}
}

@ARTICLE{2019ApJ...881L...1F,
	author = {{Fuller}, Jim and {Ma}, Linhao},
	title = "{Most Black Holes Are Born Very Slowly Rotating}",
	journal = {\apjl},
	year = 2019,
	month = aug,
	volume = {881},
	number = {1},
	eid = {L1},
	pages = {L1},
	doi = {10.3847/2041-8213/ab339b},
	archivePrefix = {arXiv},
	eprint = {1907.03714},
	primaryClass = {astro-ph.SR},
	adsurl = {https://ui.adsabs.harvard.edu/abs/2019ApJ...881L...1F}
}

@ARTICLE{2002A&A...381..923S,
	author = {{Spruit}, H.~C.},
	title = "{Dynamo action by differential rotation in a stably stratified stellar interior}",
	journal = {\aap},
	year = 2002,
	month = jan,
	volume = {381},
	pages = {923-932},
	doi = {10.1051/0004-6361:20011465},
	archivePrefix = {arXiv},
	eprint = {astro-ph/0108207},
	primaryClass = {astro-ph},
	adsurl = {https://ui.adsabs.harvard.edu/abs/2002A&A...381..923S}
}

@ARTICLE{1992MNRAS.258..811P,
	author = {{Pringle}, J.~E.},
	title = "{A simple approach to the evolution of twisted accretion discs}",
	journal = {\mnras},
	year = 1992,
	month = oct,
	volume = {258},
	number = {4},
	pages = {811-818},
	doi = {10.1093/mnras/258.4.811},
	adsurl = {https://ui.adsabs.harvard.edu/abs/1992MNRAS.258..811P}
}

@ARTICLE{2011A&A...527A..17S,
	author = {{S{\k{a}}dowski}, A. and {Abramowicz}, M. and {Bursa}, M. and {Klu{\'z}niak}, W. and {Lasota}, J.-P. and {R{\'o}{\.z}a{\'n}ska}, A.},
	title = "{Relativistic slim disks with vertical structure}",
	journal = {\aap},
	year = 2011,
	month = mar,
	volume = {527},
	eid = {A17},
	pages = {A17},
	doi = {10.1051/0004-6361/201015256},
	archivePrefix = {arXiv},
	eprint = {1006.4309},
	primaryClass = {astro-ph.HE},
	adsurl = {https://ui.adsabs.harvard.edu/abs/2011A&A...527A..17S}
}

@ARTICLE{2009ApJS..183..171S,
	author = {{S{\k{a}}dowski}, Aleksander},
	title = "{Slim Disks Around Kerr Black Holes Revisited}",
	journal = {\apjs},
	year = 2009,
	month = aug,
	volume = {183},
	number = {2},
	pages = {171-178},
	doi = {10.1088/0067-0049/183/2/171},
	archivePrefix = {arXiv},
	eprint = {0906.0355},
	primaryClass = {astro-ph.HE},
	adsurl = {https://ui.adsabs.harvard.edu/abs/2009ApJS..183..171S}
}

@ARTICLE{2004ApJ...602..312G,
	author = {{Gammie}, Charles F. and {Shapiro}, Stuart L. and {McKinney}, Jonathan C.},
	title = "{Black Hole Spin Evolution}",
	journal = {\apj},
	year = 2004,
	month = feb,
	volume = {602},
	number = {1},
	pages = {312-319},
	doi = {10.1086/380996},
	archivePrefix = {arXiv},
	eprint = {astro-ph/0310886},
	primaryClass = {astro-ph},
	adsurl = {https://ui.adsabs.harvard.edu/abs/2004ApJ...602..312G}
}

@ARTICLE{2016MNRAS.459.3738I,
	author = {{Inayoshi}, Kohei and {Haiman}, Zolt{\'a}n and {Ostriker}, Jeremiah P.},
	title = "{Hyper-Eddington accretion flows on to massive black holes}",
	journal = {\mnras},
	year = 2016,
	month = jul,
	volume = {459},
	number = {4},
	pages = {3738-3755},
	doi = {10.1093/mnras/stw836},
	archivePrefix = {arXiv},
	eprint = {1511.02116},
	primaryClass = {astro-ph.HE},
	adsurl = {https://ui.adsabs.harvard.edu/abs/2016MNRAS.459.3738I}
}

@ARTICLE{1999ApJ...513..252O,
	author = {{Ostriker}, Eve C.},
	title = "{Dynamical Friction in a Gaseous Medium}",
	journal = {\apj},
	year = "1999",
	month = "Mar",
	volume = {513},
	number = {1},
	pages = {252-258},
	doi = {10.1086/306858},
	archivePrefix = {arXiv},
	eprint = {astro-ph/9810324},
	primaryClass = {astro-ph},
	adsurl = {https://ui.adsabs.harvard.edu/abs/1999ApJ...513..252O}
}

@BOOK{2002apa..book.....F,
	author = {{Frank}, Juhan and {King}, Andrew and {Raine}, Derek J.},
	title = "{Accretion Power in Astrophysics: Third Edition}",
	year = 2002,
	publisher = {Cambridge University Press},
	adsurl = {https://ui.adsabs.harvard.edu/abs/2002apa..book.....F}
}

@ARTICLE{2001MNRAS.324..573B,
	author = {{Bonnell}, I.~A. and {Clarke}, C.~J. and {Bate}, M.~R. and
	{Pringle}, J.~E.},
	title = "{Accretion in stellar clusters and the initial mass function}",
	journal = {\mnras},
	year = "2001",
	month = "Jun",
	volume = {324},
	number = {3},
	pages = {573-579},
	doi = {10.1046/j.1365-8711.2001.04311.x},
	archivePrefix = {arXiv},
	eprint = {astro-ph/0102121},
	primaryClass = {astro-ph},
	adsurl = {https://ui.adsabs.harvard.edu/abs/2001MNRAS.324..573B}
}

@ARTICLE{1971ARA&A...9..183P,
	author = {{Paczy{\'n}ski}, B.},
	title = "{Evolutionary Processes in Close Binary Systems}",
	journal = {\araa},
	year = "1971",
	month = "Jan",
	volume = {9},
	pages = {183},
	doi = {10.1146/annurev.aa.09.090171.001151},
	adsurl = {https://ui.adsabs.harvard.edu/abs/1971ARA&A...9..183P}
}

@ARTICLE{1973A&A....24..337S,
	author = {{Shakura}, N.~I. and {Sunyaev}, R.~A.},
	title = "{Black holes in binary systems. Observational appearance.}",
	journal = {\aap},
	year = 1973,
	month = jan,
	volume = {24},
	pages = {337-355},
	adsurl = {https://ui.adsabs.harvard.edu/abs/1973A&A....24..337S}
}

@ARTICLE{2011A&A...532A..41S,
	author = {{S{\k{a}}dowski}, A. and {Bursa}, M. and {Abramowicz}, M. and {Klu{\'z}niak}, W. and {Lasota}, J.-P. and {Moderski}, R. and {Safarzadeh}, M.},
	title = "{Spinning up black holes with super-critical accretion flows}",
	journal = {\aap},
	year = 2011,
	month = aug,
	volume = {532},
	eid = {A41},
	pages = {A41},
	doi = {10.1051/0004-6361/201116702},
	archivePrefix = {arXiv},
	eprint = {1102.2456},
	primaryClass = {astro-ph.HE},
	adsurl = {https://ui.adsabs.harvard.edu/abs/2011A&A...532A..41S}
}

@ARTICLE{1972ApJ...178..347B,
	author = {{Bardeen  }, James M. and {Press}, William H. and {Teukolsky}, Saul A.},
	title = "{Rotating Black Holes: Locally Nonrotating Frames, Energy Extraction, and Scalar Synchrotron Radiation}",
	journal = {\apj},
	year = 1972,
	month = dec,
	volume = {178},
	pages = {347-370},
	doi = {10.1086/151796},
	adsurl = {https://ui.adsabs.harvard.edu/abs/1972ApJ...178..347B}
}

@ARTICLE{2024Natur.632..513A,
	author = {{Adamo}, Angela and {Bradley}, Larry D. and {Vanzella}, Eros and {Claeyssens}, Ad{\'e}la{\"\i}de and {Welch}, Brian and {Diego}, Jose M. and {Mahler}, Guillaume and {Oguri}, Masamune and {Sharon}, Keren and {Abdurro'uf} and {Hsiao}, Tiger Yu-Yang and {Xu}, Xinfeng and {Messa}, Matteo and {Lassen}, Augusto E. and {Zackrisson}, Erik and {Brammer}, Gabriel and {Coe}, Dan and {Kokorev}, Vasily and {Ricotti}, Massimo and {Zitrin}, Adi and {Fujimoto}, Seiji and {Inoue}, Akio K. and {Resseguier}, Tom and {Rigby}, Jane R. and {Jim{\'e}nez-Teja}, Yolanda and {Windhorst}, Rogier A. and {Hashimoto}, Takuya and {Tamura}, Yoichi},
	title = "{Bound star clusters observed in a lensed galaxy 460 Myr after the Big Bang}",
	journal = {\nat},
	year = 2024,
	month = aug,
	volume = {632},
	number = {8025},
	pages = {513-516},
	doi = {10.1038/s41586-024-07703-7},
	archivePrefix = {arXiv},
	eprint = {2401.03224},
	primaryClass = {astro-ph.GA},
	adsurl = {https://ui.adsabs.harvard.edu/abs/2024Natur.632..513A}
}

@ARTICLE{2003ARA&A..41...57L,
	author = {{Lada}, Charles J. and {Lada}, Elizabeth A.},
	title = "{Embedded Clusters in Molecular Clouds}",
	journal = {\araa},
	year = 2003,
	month = jan,
	volume = {41},
	pages = {57-115},
	doi = {10.1146/annurev.astro.41.011802.094844},
	archivePrefix = {arXiv},
	eprint = {astro-ph/0301540},
	primaryClass = {astro-ph},
	adsurl = {https://ui.adsabs.harvard.edu/abs/2003ARA&A..41...57L}
}

@ARTICLE{1974ApJ...191..507T,
	author = {{Thorne}, Kip S.},
	title = "{Disk-Accretion onto a Black Hole. II. Evolution of the Hole}",
	journal = {\apj},
	year = 1974,
	month = jul,
	volume = {191},
	pages = {507-520},
	doi = {10.1086/152991},
	adsurl = {https://ui.adsabs.harvard.edu/abs/1974ApJ...191..507T}
}

@ARTICLE{2010MNRAS.408..752P,
	author = {{Penna}, Robert F. and {McKinney}, Jonathan C. and {Narayan}, Ramesh and {Tchekhovskoy}, Alexander and {Shafee}, Rebecca and {McClintock}, Jeffrey E.},
	title = "{Simulations of magnetized discs around black holes: effects of black hole spin, disc thickness and magnetic field geometry}",
	journal = {\mnras},
	year = 2010,
	month = oct,
	volume = {408},
	number = {2},
	pages = {752-782},
	doi = {10.1111/j.1365-2966.2010.17170.x},
	archivePrefix = {arXiv},
	eprint = {1003.0966},
	primaryClass = {astro-ph.HE},
	adsurl = {https://ui.adsabs.harvard.edu/abs/2010MNRAS.408..752P}
}

@ARTICLE{2007MNRAS.381.1617M,
	author = {{Martin}, Rebecca G. and {Pringle}, J.~E. and {Tout}, Christopher A.},
	title = "{Alignment and precession of a black hole with a warped accretion disc}",
	journal = {\mnras},
	year = 2007,
	month = nov,
	volume = {381},
	number = {4},
	pages = {1617-1624},
	doi = {10.1111/j.1365-2966.2007.12349.x},
	archivePrefix = {arXiv},
	eprint = {0708.2034},
	primaryClass = {astro-ph},
	adsurl = {https://ui.adsabs.harvard.edu/abs/2007MNRAS.381.1617M}
}

@ARTICLE{2005MNRAS.363...49K,
	author = {{King}, A.~R. and {Lubow}, S.~H. and {Ogilvie}, G.~I. and {Pringle}, J.~E.},
	title = "{Aligning spinning black holes and accretion discs}",
	journal = {\mnras},
	year = 2005,
	month = oct,
	volume = {363},
	number = {1},
	pages = {49-56},
	doi = {10.1111/j.1365-2966.2005.09378.x},
	archivePrefix = {arXiv},
	eprint = {astro-ph/0507098},
	primaryClass = {astro-ph},
	adsurl = {https://ui.adsabs.harvard.edu/abs/2005MNRAS.363...49K}
}

@ARTICLE{2015MNRAS.454.1658K,
	author = {{Kruijssen}, J.~M. Diederik},
	title = "{Globular clusters as the relics of regular star formation in `normal' high-redshift galaxies}",
	journal = {\mnras},
	year = 2015,
	month = dec,
	volume = {454},
	number = {2},
	pages = {1658-1686},
	doi = {10.1093/mnras/stv2026},
	archivePrefix = {arXiv},
	eprint = {1509.02163},
	primaryClass = {astro-ph.GA},
	adsurl = {https://ui.adsabs.harvard.edu/abs/2015MNRAS.454.1658K}
}

@ARTICLE{2017MNRAS.469L..63R,
	author = {{Renzini}, Alvio},
	title = "{Finding forming globular clusters at high redshifts}",
	journal = {\mnras},
	year = 2017,
	month = jul,
	volume = {469},
	number = {1},
	pages = {L63-L67},
	doi = {10.1093/mnrasl/slx057},
	archivePrefix = {arXiv},
	eprint = {1704.04883},
	primaryClass = {astro-ph.GA},
	adsurl = {https://ui.adsabs.harvard.edu/abs/2017MNRAS.469L..63R}
}

@ARTICLE{2018ApJ...869..119E,
	author = {{Elmegreen}, Bruce G.},
	title = "{Two Thresholds for Globular Cluster Formation and the Common Occurrence of Massive Clusters in the Early Universe}",
	journal = {\apj},
	year = 2018,
	month = dec,
	volume = {869},
	number = {2},
	eid = {119},
	pages = {119},
	doi = {10.3847/1538-4357/aaed45},
	archivePrefix = {arXiv},
	eprint = {1810.12940},
	primaryClass = {astro-ph.GA},
	adsurl = {https://ui.adsabs.harvard.edu/abs/2018ApJ...869..119E}
}

@ARTICLE{2016A&A...594A..97B,
	author = {{Belczynski}, K. and {Heger}, A. and {Gladysz}, W. and {Ruiter}, A.~J. and {Woosley}, S. and {Wiktorowicz}, G. and {Chen}, H. -Y. and {Bulik}, T. and {O'Shaughnessy}, R. and {Holz}, D.~E. and {Fryer}, C.~L. and {Berti}, E.},
	title = "{The effect of pair-instability mass loss on black-hole mergers}",
	journal = {\aap},
	year = 2016,
	month = oct,
	volume = {594},
	eid = {A97},
	pages = {A97},
	doi = {10.1051/0004-6361/201628980},
	archivePrefix = {arXiv},
	eprint = {1607.03116},
	primaryClass = {astro-ph.HE},
	adsurl = {https://ui.adsabs.harvard.edu/abs/2016A&A...594A..97B}
}

@ARTICLE{2017MNRAS.470.4739S,
	author = {{Spera}, Mario and {Mapelli}, Michela},
	title = "{Very massive stars, pair-instability supernovae and intermediate-mass black holes with the sevn code}",
	journal = {\mnras},
	year = 2017,
	month = oct,
	volume = {470},
	number = {4},
	pages = {4739-4749},
	doi = {10.1093/mnras/stx1576},
	archivePrefix = {arXiv},
	eprint = {1706.06109},
	primaryClass = {astro-ph.SR},
	adsurl = {https://ui.adsabs.harvard.edu/abs/2017MNRAS.470.4739S}
}

@ARTICLE{2021ApJ...912L..31W,
	author = {{Woosley}, S.~E. and {Heger}, Alexander},
	title = "{The Pair-instability Mass Gap for Black Holes}",
	journal = {\apjl},
	year = 2021,
	month = may,
	volume = {912},
	number = {2},
	eid = {L31},
	pages = {L31},
	doi = {10.3847/2041-8213/abf2c4},
	archivePrefix = {arXiv},
	eprint = {2103.07933},
	primaryClass = {astro-ph.SR},
	adsurl = {https://ui.adsabs.harvard.edu/abs/2021ApJ...912L..31W}
}

@ARTICLE{2014MNRAS.441..919L,
	author = {{Leigh}, N.~W.~C. and {Mastrobuono-Battisti}, A. and {Perets}, H.~B. and 
	{B{\"o}ker}, T.},
	title = "{Stellar dynamics in gas: the role of gas damping}",
	journal = {\mnras},
	archivePrefix = "arXiv",
	eprint = {1404.0379},
	primaryClass = "astro-ph.SR",
	year = 2014,
	month = jun,
	volume = 441,
	pages = {919-932},
	doi = {10.1093/mnras/stu622},
	adsurl = {http://adsabs.harvard.edu/abs/2014MNRAS.441..919L}
}

@ARTICLE{2017PhRvD..95l4046G,
	author = {{Gerosa}, Davide and {Berti}, Emanuele},
	title = "{Are merging black holes born from stellar collapse or previous mergers?}",
	journal = {\prd},
	year = 2017,
	month = jun,
	volume = {95},
	number = {12},
	eid = {124046},
	pages = {124046},
	doi = {10.1103/PhysRevD.95.124046},
	archivePrefix = {arXiv},
	eprint = {1703.06223},
	primaryClass = {gr-qc},
	adsurl = {https://ui.adsabs.harvard.edu/abs/2017PhRvD..95l4046G}
}

@ARTICLE{2019A&A...632L...8R,
	author = {{Roupas}, Zacharias and {Kazanas}, Demosthenes},
	title = "{Generation of massive stellar black holes by rapid gas accretion in primordial dense clusters}",
	journal = {\aap},
	year = 2019,
	month = dec,
	volume = {632},
	eid = {L8},
	pages = {L8},
	doi = {10.1051/0004-6361/201937002},
	adsurl = {https://ui.adsabs.harvard.edu/abs/2019A&A...632L...8R}
}

@ARTICLE{2007MNRAS.374..515L,
	author = {{Levin}, Yuri},
	title = "{Starbursts near supermassive black holes: young stars in the Galactic Centre, and gravitational waves in LISA band}",
	journal = {\mnras},
	year = 2007,
	month = jan,
	volume = {374},
	number = {2},
	pages = {515-524},
	doi = {10.1111/j.1365-2966.2006.11155.x},
	archivePrefix = {arXiv},
	eprint = {astro-ph/0603583},
	primaryClass = {astro-ph},
	adsurl = {https://ui.adsabs.harvard.edu/abs/2007MNRAS.374..515L}
}

@ARTICLE{2012MNRAS.425..460M,
	author = {{McKernan}, B. and {Ford}, K.~E.~S. and {Lyra}, W. and {Perets}, H.~B.},
	title = "{Intermediate mass black holes in AGN discs - I. Production and growth}",
	journal = {\mnras},
	year = 2012,
	month = sep,
	volume = {425},
	number = {1},
	pages = {460-469},
	doi = {10.1111/j.1365-2966.2012.21486.x},
	archivePrefix = {arXiv},
	eprint = {1206.2309},
	primaryClass = {astro-ph.GA},
	adsurl = {https://ui.adsabs.harvard.edu/abs/2012MNRAS.425..460M}
}

@ARTICLE{2018ApJ...859L..25Y,
	author = {{Yi}, Shu-Xu and {Cheng}, K.~S. and {Taam}, Ronald E.},
	title = "{The Growth of Stellar Mass Black Hole Binaries Trapped in the Accretion Disks of Active Galactic Nuclei}",
	journal = {\apjl},
	year = 2018,
	month = jun,
	volume = {859},
	number = {2},
	eid = {L25},
	pages = {L25},
	doi = {10.3847/2041-8213/aac649},
	archivePrefix = {arXiv},
	eprint = {1805.07026},
	primaryClass = {astro-ph.HE},
	adsurl = {https://ui.adsabs.harvard.edu/abs/2018ApJ...859L..25Y}
}

@ARTICLE{2020ApJ...901L..34Y,
	author = {{Yang}, Y. and {Gayathri}, V. and {Bartos}, I. and {Haiman}, Z. and {Safarzadeh}, M. and {Tagawa}, H.},
	title = "{Black Hole Formation in the Lower Mass Gap through Mergers and Accretion in AGN Disks}",
	journal = {\apjl},
	year = 2020,
	month = oct,
	volume = {901},
	number = {2},
	eid = {L34},
	pages = {L34},
	doi = {10.3847/2041-8213/abb940},
	archivePrefix = {arXiv},
	eprint = {2007.04781},
	primaryClass = {astro-ph.HE},
	adsurl = {https://ui.adsabs.harvard.edu/abs/2020ApJ...901L..34Y}
}

@ARTICLE{2019PhRvD.100d1301G,
	author = {{Gerosa}, Davide and {Berti}, Emanuele},
	title = "{Escape speed of stellar clusters from multiple-generation black-hole mergers in the upper mass gap}",
	journal = {\prd},
	year = 2019,
	month = aug,
	volume = {100},
	number = {4},
	eid = {041301},
	pages = {041301},
	doi = {10.1103/PhysRevD.100.041301},
	archivePrefix = {arXiv},
	eprint = {1906.05295},
	primaryClass = {astro-ph.HE},
	adsurl = {https://ui.adsabs.harvard.edu/abs/2019PhRvD.100d1301G}
}

@ARTICLE{2013LRR....16....4B,
	author = {{Benacquista}, Matthew J. and {Downing}, Jonathan M.~B.},
	title = "{Relativistic Binaries in Globular Clusters}",
	journal = {Living Reviews in Relativity},
	year = 2013,
	month = dec,
	volume = {16},
	number = {1},
	eid = {4},
	pages = {4},
	doi = {10.12942/lrr-2013-4},
	archivePrefix = {arXiv},
	eprint = {1110.4423},
	primaryClass = {astro-ph.SR},
	adsurl = {https://ui.adsabs.harvard.edu/abs/2013LRR....16....4B}
}

@ARTICLE{2016PhRvD..93h4029R,
	author = {{Rodriguez}, Carl L. and {Chatterjee}, Sourav and {Rasio}, Frederic A.},
	title = "{Binary black hole mergers from globular clusters: Masses, merger rates, and the impact of stellar evolution}",
	journal = {\prd},
	year = 2016,
	month = apr,
	volume = {93},
	number = {8},
	eid = {084029},
	pages = {084029},
	doi = {10.1103/PhysRevD.93.084029},
	archivePrefix = {arXiv},
	eprint = {1602.02444},
	primaryClass = {astro-ph.HE},
	adsurl = {https://ui.adsabs.harvard.edu/abs/2016PhRvD..93h4029R}
}

@ARTICLE{2019MNRAS.486.5008A,
	author = {{Antonini}, Fabio and {Gieles}, Mark and {Gualandris}, Alessia},
	title = "{Black hole growth through hierarchical black hole mergers in dense star clusters: implications for gravitational wave detections}",
	journal = {\mnras},
	year = 2019,
	month = jul,
	volume = {486},
	number = {4},
	pages = {5008-5021},
	doi = {10.1093/mnras/stz1149},
	archivePrefix = {arXiv},
	eprint = {1811.03640},
	primaryClass = {astro-ph.HE},
	adsurl = {https://ui.adsabs.harvard.edu/abs/2019MNRAS.486.5008A}
}

@ARTICLE{2020PhRvD.102d3002B,
	author = {{Baibhav}, Vishal and {Gerosa}, Davide and {Berti}, Emanuele and {Wong}, Kaze W.~K. and {Helfer}, Thomas and {Mould}, Matthew},
	title = "{The mass gap, the spin gap, and the origin of merging binary black holes}",
	journal = {\prd},
	year = 2020,
	month = aug,
	volume = {102},
	number = {4},
	eid = {043002},
	pages = {043002},
	doi = {10.1103/PhysRevD.102.043002},
	archivePrefix = {arXiv},
	eprint = {2004.00650},
	primaryClass = {astro-ph.HE},
	adsurl = {https://ui.adsabs.harvard.edu/abs/2020PhRvD.102d3002B}
}

@ARTICLE{2020ApJ...904L..13R,
	author = {{Renzo}, M. and {Cantiello}, M. and {Metzger}, B.~D. and {Jiang}, Y. -F.},
	title = "{The Stellar Merger Scenario for Black Holes in the Pair-instability Gap}",
	journal = {\apjl},
	year = 2020,
	month = dec,
	volume = {904},
	number = {2},
	eid = {L13},
	pages = {L13},
	doi = {10.3847/2041-8213/abc6a6},
	archivePrefix = {arXiv},
	eprint = {2010.00705},
	primaryClass = {astro-ph.SR},
	adsurl = {https://ui.adsabs.harvard.edu/abs/2020ApJ...904L..13R}
}

@article{Kimball_2020,
	doi = {10.3847/1538-4357/aba518},
	url = {https://dx.doi.org/10.3847/1538-4357/aba518},
	year = {2020},
	month = {sep},
	publisher = {The American Astronomical Society},
	volume = {900},
	number = {2},
	pages = {177},
	author = {Kimball, Chase and Talbot, Colm and L. Berry, Christopher P. and Carney, Matthew and Zevin, Michael and Thrane, Eric and Kalogera, Vicky},
	title = {Black Hole Genealogy: Identifying Hierarchical Mergers with Gravitational Waves},
	journal = {The Astrophysical Journal},
}

@ARTICLE{2017ApJ...840L..24F,
	author = {{Fishbach}, Maya and {Holz}, Daniel E. and {Farr}, Ben},
	title = "{Are LIGO's Black Holes Made from Smaller Black Holes?}",
	journal = {\apjl},
	year = 2017,
	month = may,
	volume = {840},
	number = {2},
	eid = {L24},
	pages = {L24},
	doi = {10.3847/2041-8213/aa7045},
	archivePrefix = {arXiv},
	eprint = {1703.06869},
	primaryClass = {astro-ph.HE},
	adsurl = {https://ui.adsabs.harvard.edu/abs/2017ApJ...840L..24F}
}

@ARTICLE{2020ApJ...903L..21S,
	author = {{Safarzadeh}, Mohammadtaher and {Haiman}, Zolt{\'a}n},
	title = "{Formation of GW190521 via Gas Accretion onto Population III Stellar Black Hole Remnants Born in High-redshift Minihalos}",
	journal = {\apjl},
	year = 2020,
	month = nov,
	volume = {903},
	number = {1},
	eid = {L21},
	pages = {L21},
	doi = {10.3847/2041-8213/abc253},
	archivePrefix = {arXiv},
	eprint = {2009.09320},
	primaryClass = {astro-ph.HE},
	adsurl = {https://ui.adsabs.harvard.edu/abs/2020ApJ...903L..21S}
}

@ARTICLE{2021MNRAS.505..339M,
	author = {{Mapelli}, Michela and {Dall'Amico}, Marco and {Bouffanais}, Yann and {Giacobbo}, Nicola and {Arca Sedda}, Manuel and {Artale}, M. Celeste and {Ballone}, Alessandro and {Di Carlo}, Ugo N. and {Iorio}, Giuliano and {Santoliquido}, Filippo and {Torniamenti}, Stefano},
	title = "{Hierarchical black hole mergers in young, globular and nuclear star clusters: the effect of metallicity, spin and cluster properties}",
	journal = {\mnras},
	year = 2021,
	month = jul,
	volume = {505},
	number = {1},
	pages = {339-358},
	doi = {10.1093/mnras/stab1334},
	archivePrefix = {arXiv},
	eprint = {2103.05016},
	primaryClass = {astro-ph.HE},
	adsurl = {https://ui.adsabs.harvard.edu/abs/2021MNRAS.505..339M}
}

@ARTICLE{2022MNRAS.512..884R,
	author = {{Rizzuto}, Francesco Paolo and {Naab}, Thorsten and {Spurzem}, Rainer and {Arca-Sedda}, Manuel and {Giersz}, Mirek and {Ostriker}, Jeremiah Paul and {Banerjee}, Sambaran},
	title = "{Black hole mergers in compact star clusters and massive black hole formation beyond the mass gap}",
	journal = {\mnras},
	year = 2022,
	month = may,
	volume = {512},
	number = {1},
	pages = {884-898},
	doi = {10.1093/mnras/stac231},
	archivePrefix = {arXiv},
	eprint = {2108.11457},
	primaryClass = {astro-ph.GA},
	adsurl = {https://ui.adsabs.harvard.edu/abs/2022MNRAS.512..884R}
}

@ARTICLE{2023MNRAS.526..429A,
	author = {{Arca Sedda}, Manuel and {Kamlah}, Albrecht W.~H. and {Spurzem}, Rainer and {Rizzuto}, Francesco Paolo and {Naab}, Thorsten and {Giersz}, Mirek and {Berczik}, Peter},
	title = "{The DRAGON-II simulations - II. Formation mechanisms, mass, and spin of intermediate-mass black holes in star clusters with up to 1 million stars}",
	journal = {\mnras},
	year = 2023,
	month = nov,
	volume = {526},
	number = {1},
	pages = {429-442},
	doi = {10.1093/mnras/stad2292},
	archivePrefix = {arXiv},
	eprint = {2307.04806},
	primaryClass = {astro-ph.GA},
	adsurl = {https://ui.adsabs.harvard.edu/abs/2023MNRAS.526..429A}
}

@ARTICLE{2024A&A...688A.148T,
	author = {{Torniamenti}, Stefano and {Mapelli}, Michela and {P{\'e}rigois}, Carole and {Arca Sedda}, Manuel and {Artale}, Maria Celeste and {Dall'Amico}, Marco and {Vaccaro}, Maria Paola},
	title = "{Hierarchical binary black hole mergers in globular clusters: Mass function and evolution with redshift}",
	journal = {\aap},
	year = 2024,
	month = aug,
	volume = {688},
	eid = {A148},
	pages = {A148},
	doi = {10.1051/0004-6361/202449272},
	archivePrefix = {arXiv},
	eprint = {2401.14837},
	primaryClass = {astro-ph.HE},
	adsurl = {https://ui.adsabs.harvard.edu/abs/2024A&A...688A.148T}
}

@ARTICLE{2024Sci...384.1488F,
	author = {{Fujii}, Michiko S. and {Wang}, Long and {Tanikawa}, Ataru and {Hirai}, Yutaka and {Saitoh}, Takayuki R.},
	title = "{Simulations predict intermediate-mass black hole formation in globular clusters}",
	journal = {Science},
	year = 2024,
	month = jun,
	volume = {384},
	number = {6703},
	pages = {1488-1492},
	doi = {10.1126/science.adi4211},
	archivePrefix = {arXiv},
	eprint = {2406.06772},
	primaryClass = {astro-ph.GA},
	adsurl = {https://ui.adsabs.harvard.edu/abs/2024Sci...384.1488F}
}

@ARTICLE{2024AJ....167..191P,
	author = {{Purohit}, Rujuta A. and {Fragione}, Giacomo and {Rasio}, Frederic A. and {Petter}, Grayson C. and {Hickox}, Ryan C.},
	title = "{Binary Black Hole Mergers and Intermediate-mass Black Holes in Dense Star Clusters with Collisional Runaways}",
	journal = {\aj},
	year = 2024,
	month = may,
	volume = {167},
	number = {5},
	eid = {191},
	pages = {191},
	doi = {10.3847/1538-3881/ad3103},
	archivePrefix = {arXiv},
	eprint = {2401.06731},
	primaryClass = {astro-ph.GA},
	adsurl = {https://ui.adsabs.harvard.edu/abs/2024AJ....167..191P}
}

@ARTICLE{2025PhRvD.111f3039B,
	author = {{Baumgarte}, Thomas W. and {Shapiro}, Stuart L.},
	title = "{Boosting the growth of intermediate-mass black holes: Collisions with massive stars}",
	journal = {\prd},
	year = 2025,
	month = mar,
	volume = {111},
	number = {6},
	eid = {063039},
	pages = {063039},
	doi = {10.1103/PhysRevD.111.063039},
	archivePrefix = {arXiv},
	eprint = {2502.14955},
	primaryClass = {astro-ph.HE},
	adsurl = {https://ui.adsabs.harvard.edu/abs/2025PhRvD.111f3039B}
}

@ARTICLE{2025ApJ...988...15L,
	author = {{Lee}, Seungjae and {Lee}, Hyung Mok and {Kim}, Ji-hoon and {Spurzem}, Rainer and {Hong}, Jongsuk and {Chung}, Eunwoo},
	title = "{Formation and Evolution of Compact Binaries Containing Intermediate-mass Black Holes in Dense Star Clusters}",
	journal = {\apj},
	year = 2025,
	month = jul,
	volume = {988},
	number = {1},
	eid = {15},
	pages = {15},
	doi = {10.3847/1538-4357/adde52},
	archivePrefix = {arXiv},
	eprint = {2503.22109},
	primaryClass = {astro-ph.GA},
	adsurl = {https://ui.adsabs.harvard.edu/abs/2025ApJ...988...15L}
}

@ARTICLE{2025ApJ...994L..37K,
	author = {{K{\i}ro{\u{g}}lu}, Fulya and {Kremer}, Kyle and {Rasio}, Frederic A.},
	title = "{Beyond Hierarchical Mergers: Accretion-driven Origins of Massive, Highly Spinning Black Holes in Dense Star Clusters}",
	journal = {\apjl},
	year = 2025,
	month = dec,
	volume = {994},
	number = {2},
	eid = {L37},
	pages = {L37},
	doi = {10.3847/2041-8213/ae1eeb},
	archivePrefix = {arXiv},
	eprint = {2509.05415},
	primaryClass = {astro-ph.HE},
	adsurl = {https://ui.adsabs.harvard.edu/abs/2025ApJ...994L..37K}
}

@ARTICLE{2025arXiv250808558B,
	author = {{Bartos}, Imre and {Haiman}, Zoltan},
	title = "{Accretion is All You Need: Black Hole Spin Alignment in Merger GW231123 Indicates Accretion Pathway}",
	journal = {arXiv e-prints},
	year = 2025,
	month = aug,
	eid = {arXiv:2508.08558},
	pages = {arXiv:2508.08558},
	doi = {10.48550/arXiv.2508.08558},
	archivePrefix = {arXiv},
	eprint = {2508.08558},
	primaryClass = {astro-ph.HE},
	adsurl = {https://ui.adsabs.harvard.edu/abs/2025arXiv250808558B}
}

@ARTICLE{2025ApJ...993L..54G,
	author = {{Gottlieb}, Ore and {Metzger}, Brian D. and {Issa}, Danat and {Li}, Sean E. and {Renzo}, Mathieu and {Isi}, Maximiliano},
	title = "{Spinning into the Gap: Direct-horizon Collapse as the Origin of GW231123 from End-to-end General-relativistic Magnetohydrodynamic Simulations}",
	journal = {\apjl},
	year = 2025,
	month = nov,
	volume = {993},
	number = {2},
	eid = {L54},
	pages = {L54},
	doi = {10.3847/2041-8213/ae0d81},
	archivePrefix = {arXiv},
	eprint = {2508.15887},
	primaryClass = {astro-ph.HE},
	adsurl = {https://ui.adsabs.harvard.edu/abs/2025ApJ...993L..54G}
}

@ARTICLE{2023PhRvX..13d1039A,
	author = {{Abbott}, R. and others},
	title = "{GWTC-3: Compact Binary Coalescences Observed by LIGO and Virgo during the Second Part of the Third Observing Run}",
	journal = {Physical Review X},
	year = 2023,
	month = oct,
	volume = {13},
	number = {4},
	eid = {041039},
	pages = {041039},
	doi = {10.1103/PhysRevX.13.041039},
	archivePrefix = {arXiv},
	eprint = {2111.03606},
	primaryClass = {gr-qc},
	adsurl = {https://ui.adsabs.harvard.edu/abs/2023PhRvX..13d1039A}
}

@ARTICLE{2025ApJ...993L..30L,
	author = {{Liu}, Shuai and {Wang}, Long and {Tanikawa}, Ataru and {Wu}, Weiwei and {Fujii}, Michiko S.},
	title = "{On the Formation of GW231123 in Population III Star Clusters}",
	journal = {\apjl},
	year = 2025,
	month = nov,
	volume = {993},
	number = {1},
	eid = {L30},
	pages = {L30},
	doi = {10.3847/2041-8213/ae1024},
	archivePrefix = {arXiv},
	eprint = {2510.05634},
	primaryClass = {astro-ph.GA},
	adsurl = {https://ui.adsabs.harvard.edu/abs/2025ApJ...993L..30L}
}

@BOOK{Spitzer_1987degc,
	author = {{Spitzer}, L.},
	title = "{Dynamical evolution of globular clusters}",
	booktitle = {Princeton, NJ, Princeton University Press, 1987, 191 p.},
	year = 1987,
	publisher = "Princeton University Press",
	adsurl = {http://adsabs.harvard.edu/abs/1987degc.book.....S}
}

@ARTICLE{2015MNRAS.448.1526N,
	author = {{Nealon}, Rebecca and {Price}, Daniel J. and {Nixon}, Chris J.},
	title = "{On the Bardeen-Petterson effect in black hole accretion discs}",
	journal = {\mnras},
	year = 2015,
	month = apr,
	volume = {448},
	number = {2},
	pages = {1526-1540},
	doi = {10.1093/mnras/stv014},
	archivePrefix = {arXiv},
	eprint = {1501.01687},
	primaryClass = {astro-ph.HE},
	adsurl = {https://ui.adsabs.harvard.edu/abs/2015MNRAS.448.1526N}
}

@ARTICLE{2012ApJ...757L..24N,
	author = {{Nixon}, Chris and {King}, Andrew and {Price}, Daniel and {Frank}, Juhan},
	title = "{Tearing up the Disk: How Black Holes Accrete}",
	journal = {\apjl},
	year = 2012,
	month = oct,
	volume = {757},
	number = {2},
	eid = {L24},
	pages = {L24},
	doi = {10.1088/2041-8205/757/2/L24},
	archivePrefix = {arXiv},
	eprint = {1209.1393},
	primaryClass = {astro-ph.HE},
	adsurl = {https://ui.adsabs.harvard.edu/abs/2012ApJ...757L..24N}
}

@ARTICLE{2007MNRAS.381.1287L,
	author = {{Lodato}, Giuseppe and {Pringle}, J.~E.},
	title = "{Warp diffusion in accretion discs: a numerical investigation}",
	journal = {\mnras},
	year = 2007,
	month = nov,
	volume = {381},
	number = {3},
	pages = {1287-1300},
	doi = {10.1111/j.1365-2966.2007.12332.x},
	archivePrefix = {arXiv},
	eprint = {0708.1124},
	primaryClass = {astro-ph},
	adsurl = {https://ui.adsabs.harvard.edu/abs/2007MNRAS.381.1287L}
}

@ARTICLE{2009MNRAS.399.2249P,
	author = {{Perego}, A. and {Dotti}, M. and {Colpi}, M. and {Volonteri}, M.},
	title = "{Mass and spin co-evolution during the alignment of a black hole in a warped accretion disc}",
	journal = {\mnras},
	year = 2009,
	month = nov,
	volume = {399},
	number = {4},
	pages = {2249-2263},
	doi = {10.1111/j.1365-2966.2009.15427.x},
	archivePrefix = {arXiv},
	eprint = {0907.3742},
	primaryClass = {astro-ph.CO},
	adsurl = {https://ui.adsabs.harvard.edu/abs/2009MNRAS.399.2249P}
}

@ARTICLE{1999MNRAS.309..929N,
	author = {{Nelson}, Richard P. and {Papaloizou}, John C.~B.},
	title = "{Hydrodynamic simulations of propagating WARPS and bending waves in accretion discs}",
	journal = {\mnras},
	year = 1999,
	month = nov,
	volume = {309},
	number = {4},
	pages = {929-940},
	doi = {10.1046/j.1365-8711.1999.02894.x},
	archivePrefix = {arXiv},
	eprint = {astro-ph/9907076},
	primaryClass = {astro-ph},
	adsurl = {https://ui.adsabs.harvard.edu/abs/1999MNRAS.309..929N}
}

@ARTICLE{2021ARA&A..59..117R,
	author = {{Reynolds}, Christopher S.},
	title = "{Observational Constraints on Black Hole Spin}",
	journal = {\araa},
	year = 2021,
	month = sep,
	volume = {59},
	pages = {117-154},
	doi = {10.1146/annurev-astro-112420-035022},
	archivePrefix = {arXiv},
	eprint = {2011.08948},
	primaryClass = {astro-ph.HE},
	adsurl = {https://ui.adsabs.harvard.edu/abs/2021ARA&A..59..117R}
}

@ARTICLE{2016A&A...587A..53K,
	author = {{Krause}, Martin G.~H. and {Charbonnel}, Corinne and {Bastian}, Nate and {Diehl}, Roland},
	title = "{Gas expulsion in massive star clusters?. Constraints from observations of young and gas-free objects}",
	journal = {\aap},
	year = 2016,
	month = mar,
	volume = {587},
	eid = {A53},
	pages = {A53},
	doi = {10.1051/0004-6361/201526685},
	archivePrefix = {arXiv},
	eprint = {1512.04256},
	primaryClass = {astro-ph.GA},
	adsurl = {https://ui.adsabs.harvard.edu/abs/2016A&A...587A..53K}
}

@ARTICLE{2002MNRAS.330..232C,
	author = {{Miller}, M. Coleman and {Hamilton}, Douglas P.},
	title = "{Production of intermediate-mass black holes in globular clusters}",
	journal = {\mnras},
	year = 2002,
	month = feb,
	volume = {330},
	number = {1},
	pages = {232-240},
	doi = {10.1046/j.1365-8711.2002.05112.x},
	archivePrefix = {arXiv},
	eprint = {astro-ph/0106188},
	primaryClass = {astro-ph},
	adsurl = {https://ui.adsabs.harvard.edu/abs/2002MNRAS.330..232C}
}

@ARTICLE{2006ApJ...637..937O,
	author = {{O'Leary}, Ryan M. and {Rasio}, Frederic A. and {Fregeau}, John M. and {Ivanova}, Natalia and {O'Shaughnessy}, Richard},
	title = "{Binary Mergers and Growth of Black Holes in Dense Star Clusters}",
	journal = {\apj},
	year = 2006,
	month = feb,
	volume = {637},
	number = {2},
	pages = {937-951},
	doi = {10.1086/498446},
	archivePrefix = {arXiv},
	eprint = {astro-ph/0508224},
	primaryClass = {astro-ph},
	adsurl = {https://ui.adsabs.harvard.edu/abs/2006ApJ...637..937O}
}

@ARTICLE{2025ApJ...993L..25A,
	author = {{Abac}, A.~G. and {Abouelfettouh}, I. and {Acernese}, F. and {Ackley}, K. and {Adamcewicz}, C. and {Adhicary}, S. and {Adhikari}, D. and {Adhikari}, N. and {Adhikari}, R.~X. and {Adkins}, V.~K. and {Afroz}, S. and {Agapito}, A. and {Agarwal}, D. and {Agathos}, M. and {Aggarwal}, N. and {Aggarwal}, S. and {Aguiar}, O.~D. and {Ahrend}, I.-L. and {Aiello}, L. and {Ain}, A. and {Ajith}, P. and {Akutsu}, T. and {Albanesi}, S. and {Ali}, W. and {Al-Kershi}, S. and {All{\'e}n{\'e}}, C. and {Allocca}, A. and {Al-Shammari}, S. and {Altin}, P.~A. and {Alvarez-Lopez}, S. and {Amar}, W. and {Amarasinghe}, O. and {Amato}, A. and {Amicucci}, F. and {Amra}, C. and {Ananyeva}, A. and {Anderson}, S.~B. and {Anderson}, W.~G. and {Andia}, M. and {Ando}, M. and {Andr{\'e}s-Carcasona}, M. and {Andri{\'c}}, T. and {Anglin}, J. and {Ansoldi}, S. and {Antelis}, J.~M. and {Antier}, S. and {Aoumi}, M. and {Appavuravther}, E.~Z. and {Appert}, S. and {Apple}, S.~K. and {Arai}, K. and {Alvarez}, C. Araujo and {Araya}, A. and {Araya}, M.~C. and {Sedda}, M. Arca and {Areeda}, J.~S. and {Aritomi}, N. and {Armato}, F. and {Armstrong}, S. and {Arnaud}, N. and {Arogeti}, M. and {Aronson}, S.~M. and {Arun}, K.~G. and {Ashton}, G. and {Aso}, Y. and {Asprea}, L. and {Assiduo}, M. and {Assis de Souza Melo}, S. and {Aston}, S.~M. and {Astone}, P. and {Attadio}, F. and {Aubin}, F. and {AultONeal}, K. and {Avallone}, G. and {Avila}, E.~A. and {Babak}, S. and {Badger}, C. and {Bae}, S. and {Bagnasco}, S. and {Baiotti}, L. and {Bajpai}, R. and {Baka}, T. and {Baker}, A.~M. and {Baker}, K.~A. and {Baker}, T. and {Baldi}, G. and {Baldicchi}, N. and {Ball}, M. and {Ballardin}, G. and {Ballmer}, S.~W. and {Banagiri}, S. and {Banerjee}, B. and {Bankar}, D. and {Baptiste}, T.~M. and {Baral}, P. and {Baratti}, M. and {Barayoga}, J.~C. and {Barish}, B.~C. and {Barker}, D. and {Barman}, N. and {Barneo}, P. and {Barone}, F. and {Barr}, B. and {Barsotti}, L. and {Barsuglia}, M. and {Barta}, D. and {Bartoletti}, A.~M. and {Barton}, M.~A. and {Bartos}, I. and {Basalaev}, A. and {Bassiri}, R. and {Basti}, A. and {Bawaj}, M. and {Baxi}, P. and {Bayley}, J.~C. and {Baylor}, A.~C. and {Baynard}, II, P.~A. and {Bazzan}, M. and {Bedakihale}, V.~M. and {Beirnaert}, F. and {Bejger}, M. and {Belardinelli}, D. and {Bell}, A.~S. and {Bellie}, D.~S. and {Bellizzi}, L. and {Benoit}, W. and {Bentara}, I. and {Bentley}, J.~D. and {Ben Yaala}, M. and {Bera}, S. and {Bergamin}, F. and {Berger}, B.~K. and {Bernuzzi}, S. and {Beroiz}, M. and {Berry}, C.~P.~L. and {Bersanetti}, D. and {Bertheas}, T. and {Bertolini}, A. and {Betzwieser}, J. and {Beveridge}, D. and {Bevilacqua}, G. and {Bevins}, N. and {Bhandare}, R. and {Bhatt}, R. and {Bhattacharjee}, D. and {Bhattacharyya}, S. and {Bhaumik}, S. and {Bhagwat}, S. and {Biancalana}, V. and {Bianchi}, A. and {Bilenko}, I.~A. and {Billingsley}, G. and {Binetti}, A. and {Bini}, S. and {Binu}, C. and {Biot}, S. and {Birnholtz}, O. and {Biscoveanu}, S. and {Bisht}, A. and {Bitossi}, M. and {Bizouard}, M.-A. and {Blaber}, S. and {Blackburn}, J.~K. and {Blagg}, L.~A. and {Blair}, C.~D. and {Blair}, D.~G. and {Bode}, N. and {Boettner}, N. and {Boileau}, G. and {Boldrini}, M. and {Bolingbroke}, G.~N. and {Bolliand}, A. and {Bonavena}, L.~D. and {Bondarescu}, R. and {Bondu}, F. and {Bonilla}, E. and {Bonilla}, M.~S. and {Bonino}, A. and {Bonnand}, R. and {Borchers}, A. and {Borhanian}, S. and {Boschi}, V. and {Bose}, S. and {Bossilkov}, V. and {Bothra}, Y. and {Boudon}, A. and {Bourg}, L. and {Boyle}, M. and {Bozzi}, A. and {Bradaschia}, C. and {Brady}, P.~R. and {Branch}, A. and {Branchesi}, M. and {Braun}, I. and {Briant}, T. and {Brillet}, A. and {Brinkmann}, M. and {Brockill}, P. and {Brockmueller}, E. and {Brooks}, A.~F.},
	title = "{GW231123: A Binary Black Hole Merger with Total Mass 190?265 M$_{{\ensuremath{\odot}}}$}",
	journal = {\apjl},
	year = 2025,
	month = nov,
	volume = {993},
	number = {1},
	eid = {L25},
	pages = {L25},
	doi = {10.3847/2041-8213/ae0c9c},
	archivePrefix = {arXiv},
	eprint = {2507.08219},
	primaryClass = {astro-ph.HE},
	adsurl = {https://ui.adsabs.harvard.edu/abs/2025ApJ...993L..25A}
}

@ARTICLE{2019A&A...627A..24G,
	author = {{Groh}, J.~H. and {Ekstr{\"o}m}, S. and {Georgy}, C. and {Meynet}, G. and {Choplin}, A. and {Eggenberger}, P. and {Hirschi}, R. and {Maeder}, A. and {Murphy}, L.~J. and {Boian}, I. and {Farrell}, E.~J.},
	title = "{Grids of stellar models with rotation. IV. Models from 1.7 to 120 M$_{{\ensuremath{\odot}}}$ at a metallicity Z = 0.0004}",
	journal = {\aap},
	year = 2019,
	month = jul,
	volume = {627},
	eid = {A24},
	pages = {A24},
	doi = {10.1051/0004-6361/201833720},
	archivePrefix = {arXiv},
	eprint = {1904.04009},
	primaryClass = {astro-ph.SR},
	adsurl = {https://ui.adsabs.harvard.edu/abs/2019A&A...627A..24G}
}

@ARTICLE{2002ApJ...577..389K,
	author = {{Kudritzki}, Rolf P.},
	title = "{Line-driven Winds, Ionizing Fluxes, and Ultraviolet Spectra of Hot Stars at Extremely Low Metallicity. I. Very Massive O Stars}",
	journal = {\apj},
	year = 2002,
	month = sep,
	volume = {577},
	number = {1},
	pages = {389-408},
	doi = {10.1086/342178},
	archivePrefix = {arXiv},
	eprint = {astro-ph/0205210},
	primaryClass = {astro-ph},
	adsurl = {https://ui.adsabs.harvard.edu/abs/2002ApJ...577..389K}
}

@ARTICLE{2003MNRAS.340..227B,
	author = {{Baumgardt}, Holger and {Makino}, Junichiro},
	title = "{Dynamical evolution of star clusters in tidal fields}",
	journal = {\mnras},
	year = 2003,
	month = mar,
	volume = {340},
	number = {1},
	pages = {227-246},
	doi = {10.1046/j.1365-8711.2003.06286.x},
	archivePrefix = {arXiv},
	eprint = {astro-ph/0211471},
	primaryClass = {astro-ph},
	adsurl = {https://ui.adsabs.harvard.edu/abs/2003MNRAS.340..227B}
}

@ARTICLE{2008A&ARv..16..209P,
	author = {{Puls}, Joachim and {Vink}, Jorick S. and {Najarro}, Francisco},
	title = "{Mass loss from hot massive stars}",
	journal = {\aapr},
	year = 2008,
	month = dec,
	volume = {16},
	number = {3-4},
	pages = {209-325},
	doi = {10.1007/s00159-008-0015-8},
	archivePrefix = {arXiv},
	eprint = {0811.0487},
	primaryClass = {astro-ph},
	adsurl = {https://ui.adsabs.harvard.edu/abs/2008A&ARv..16..209P}
}

@ARTICLE{2007MNRAS.380.1589B,
	author = {{Baumgardt}, H. and {Kroupa}, P.},
	title = "{A comprehensive set of simulations studying the influence of gas expulsion on star cluster evolution}",
	journal = {\mnras},
	year = 2007,
	month = oct,
	volume = {380},
	number = {4},
	pages = {1589-1598},
	doi = {10.1111/j.1365-2966.2007.12209.x},
	archivePrefix = {arXiv},
	eprint = {0707.1944},
	primaryClass = {astro-ph},
	adsurl = {https://ui.adsabs.harvard.edu/abs/2007MNRAS.380.1589B}
}

@ARTICLE{2020MNRAS.499..873S,
	author = {{Sander}, Andreas A.~C. and {Vink}, Jorick S.},
	title = "{On the nature of massive helium star winds and Wolf-Rayet-type mass-loss}",
	journal = {\mnras},
	year = 2020,
	month = nov,
	volume = {499},
	number = {1},
	pages = {873-892},
	doi = {10.1093/mnras/staa2712},
	archivePrefix = {arXiv},
	eprint = {2009.01849},
	primaryClass = {astro-ph.SR},
	adsurl = {https://ui.adsabs.harvard.edu/abs/2020MNRAS.499..873S}
}

@ARTICLE{2001A&A...369..574V,
	author = {{Vink}, Jorick S. and {de Koter}, A. and {Lamers}, H.~J.~G.~L.~M.},
	title = "{Mass-loss predictions for O and B stars as a function of metallicity}",
	journal = {\aap},
	year = 2001,
	month = apr,
	volume = {369},
	pages = {574-588},
	doi = {10.1051/0004-6361:20010127},
	archivePrefix = {arXiv},
	eprint = {astro-ph/0101509},
	primaryClass = {astro-ph},
	adsurl = {https://ui.adsabs.harvard.edu/abs/2001A&A...369..574V}
}

@ARTICLE{2017A&A...603A.118R,
	author = {{Renzo}, M. and {Ott}, C.~D. and {Shore}, S.~N. and {de Mink}, S.~E.},
	title = "{Systematic survey of the effects of wind mass loss algorithms on the evolution of single massive stars}",
	journal = {\aap},
	year = 2017,
	month = jul,
	volume = {603},
	eid = {A118},
	pages = {A118},
	doi = {10.1051/0004-6361/201730698},
	archivePrefix = {arXiv},
	eprint = {1703.09705},
	primaryClass = {astro-ph.SR},
	adsurl = {https://ui.adsabs.harvard.edu/abs/2017A&A...603A.118R}
}

@ARTICLE{2013A&A...550A..49K,
	author = {{Krause}, M. and {Fierlinger}, K. and {Diehl}, R. and {Burkert}, A. and {Voss}, R. and {Ziegler}, U.},
	title = "{Feedback by massive stars and the emergence of superbubbles. I. Energy efficiency and Vishniac instabilities}",
	journal = {\aap},
	year = 2013,
	month = feb,
	volume = {550},
	eid = {A49},
	pages = {A49},
	doi = {10.1051/0004-6361/201220060},
	archivePrefix = {arXiv},
	eprint = {1207.7231},
	primaryClass = {astro-ph.GA},
	adsurl = {https://ui.adsabs.harvard.edu/abs/2013A&A...550A..49K}
}

@ARTICLE{2006MNRAS.373L..90K,
	author = {{King}, A.~R. and {Pringle}, J.~E.},
	title = "{Growing supermassive black holes by chaotic accretion}",
	journal = {\mnras},
	year = 2006,
	month = nov,
	volume = {373},
	number = {1},
	pages = {L90-L92},
	doi = {10.1111/j.1745-3933.2006.00249.x},
	archivePrefix = {arXiv},
	eprint = {astro-ph/0609598},
	primaryClass = {astro-ph},
	adsurl = {https://ui.adsabs.harvard.edu/abs/2006MNRAS.373L..90K}
}

@ARTICLE{2008MNRAS.385.1621K,
	author = {{King}, A.~R. and {Pringle}, J.~E. and {Hofmann}, J.~A.},
	title = "{The evolution of black hole mass and spin in active galactic nuclei}",
	journal = {\mnras},
	year = 2008,
	month = apr,
	volume = {385},
	number = {3},
	pages = {1621-1627},
	doi = {10.1111/j.1365-2966.2008.12943.x},
	archivePrefix = {arXiv},
	eprint = {0801.1564},
	primaryClass = {astro-ph},
	adsurl = {https://ui.adsabs.harvard.edu/abs/2008MNRAS.385.1621K}
}

@ARTICLE{2025A&A...702A.208R,
	author = {{Roupas}, Zacharias},
	title = "{Black hole mass function shift in proto-stellar clusters driven by gas accretion}",
	journal = {\aap},
	year = 2025,
	month = oct,
	volume = {702},
	eid = {A208},
	pages = {A208},
	doi = {10.1051/0004-6361/202556434},
	archivePrefix = {arXiv},
	eprint = {2509.08448},
	primaryClass = {astro-ph.GA},
	adsurl = {https://ui.adsabs.harvard.edu/abs/2025A&A...702A.208R}
}

@ARTICLE{1974Ap&SS..28...45B,
	author = {{Bisnovatyi-Kogan}, G.~S. and {Ruzmaikin}, A.~A.},
	title = "{The Accretion of Matter by a Collapsing Star in the Presence of a Magnetic Field}",
	journal = {\apss},
	year = 1974,
	month = may,
	volume = {28},
	number = {1},
	pages = {45-59},
	doi = {10.1007/BF00642237},
	adsurl = {https://ui.adsabs.harvard.edu/abs/1974Ap&SS..28...45B}
}

@ARTICLE{2011MNRAS.418L..79T,
	author = {{Tchekhovskoy}, Alexander and {Narayan}, Ramesh and {McKinney}, Jonathan C.},
	title = "{Efficient generation of jets from magnetically arrested accretion on a rapidly spinning black hole}",
	journal = {\mnras},
	year = 2011,
	month = nov,
	volume = {418},
	number = {1},
	pages = {L79-L83},
	doi = {10.1111/j.1745-3933.2011.01147.x},
	archivePrefix = {arXiv},
	eprint = {1108.0412},
	primaryClass = {astro-ph.HE},
	adsurl = {https://ui.adsabs.harvard.edu/abs/2011MNRAS.418L..79T}
}

@ARTICLE{2023ApJ...954L..22R,
	author = {{Ricarte}, Angelo and {Narayan}, Ramesh and {Curd}, Brandon},
	title = "{Recipes for Jet Feedback and Spin Evolution of Black Holes with Strongly Magnetized Super-Eddington Accretion Disks}",
	journal = {\apjl},
	year = 2023,
	month = sep,
	volume = {954},
	number = {1},
	eid = {L22},
	pages = {L22},
	doi = {10.3847/2041-8213/aceda5},
	archivePrefix = {arXiv},
	eprint = {2307.04621},
	primaryClass = {astro-ph.HE},
	adsurl = {https://ui.adsabs.harvard.edu/abs/2023ApJ...954L..22R}
}
	\bibliographystyle{aa}

	\appendix
	
	\section{Stochastic kick timesteps} \label{app:dt_stoch}
	
	Our estimate of the stochastic timestep $\Delta t_{\rm stoch}$ for applying velocity increments onto each BH relies on a coarse-grained description of the unresolved perturbations experienced by the BH as it moves through the local stellar and gaseous background. In this picture the natural reference timescale is the gravitational dynamical timescale $\tau_{\rm grav}(r_{\bullet}) = \sqrt{r_{\bullet}^3/GM_{\rm enc}(r_{\bullet})}$, while additional timescales are used as consistency bounds.
	Our code calculates the adaptive local timestep for each BH as
	\begin{equation}\label{eq:Deltat_stoch}
		\Delta t_{\rm stoch} = f_{\rm dt} \cdot \min\left\{ \tau_{\rm grav},\, \tau_{\rm df}, \, 
		\tau_{\rm cross},\, \tau_{\rm turb} \right\},
	\end{equation}
	where we denote the dynamical friction timescale $\tau_{\rm df} = 1 / (\eta_{\bullet, \star} + \eta_{\bullet,{\rm gas}})$, the core crossing time $\tau_{\rm cross} = r_{c,\star} / v_{\bullet}$, the supersonic turbulence correlation time $\tau_{\rm turb} = r_{\rm c, gas} / (2c_{s,{\rm gas}})$ and a control parameter $f_{\rm dt}$.
	The dynamical friction coefficient behaves as $\eta_{\bullet} \propto 4\pi m_{\bullet} G^2 \rho / v_{\bullet}^3$; in particular we use the standard Chandrasekhar formula for the star component and the prescription of \citet{1999ApJ...513..252O} for the 
	gas component (see \citetalias{2025A&A...702A.208R} for details). Here $c_{s,{\rm gas}}$ denotes the sound speed of the ambient cluster gas, not of the accretion disk gas. 
	
\begin{figure}[h!]
		\centering
		\begin{subfigure}{\columnwidth}
			\centering
			\includegraphics[width=0.9\columnwidth]{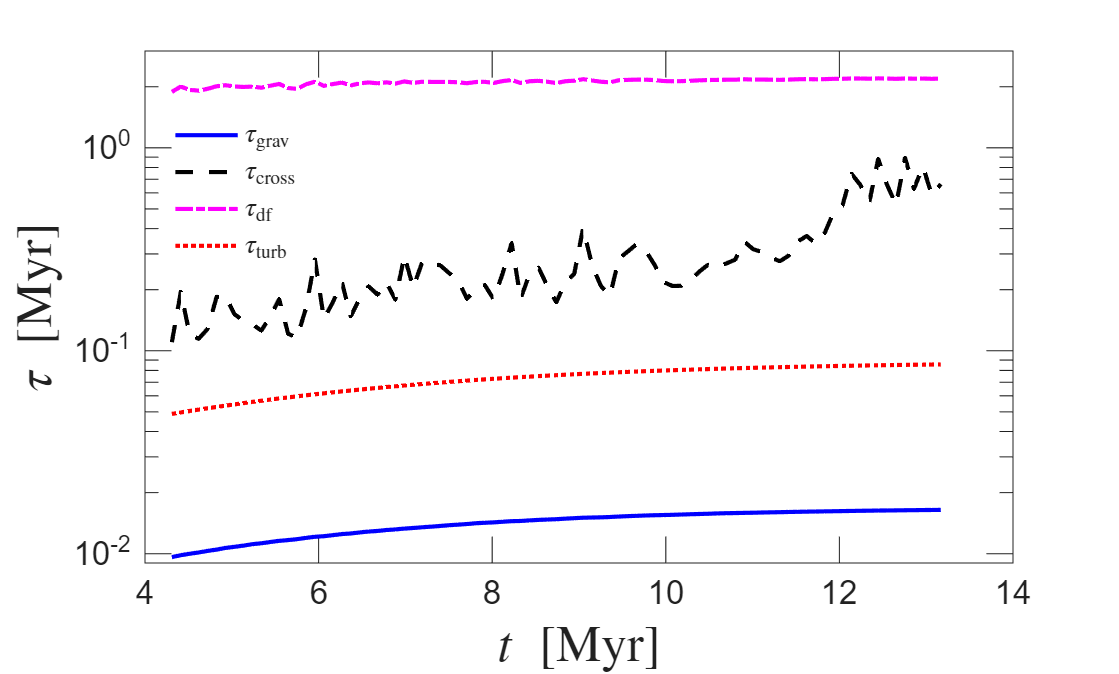}
			\caption{$m_{\bullet,{\rm f}} = 100\,{\rm M}_{\odot}$}
			\label{fig:stoch_timescales_t_m100}
		\end{subfigure}
		\begin{subfigure}{\columnwidth}
			\centering
			\includegraphics[width=0.9\columnwidth]{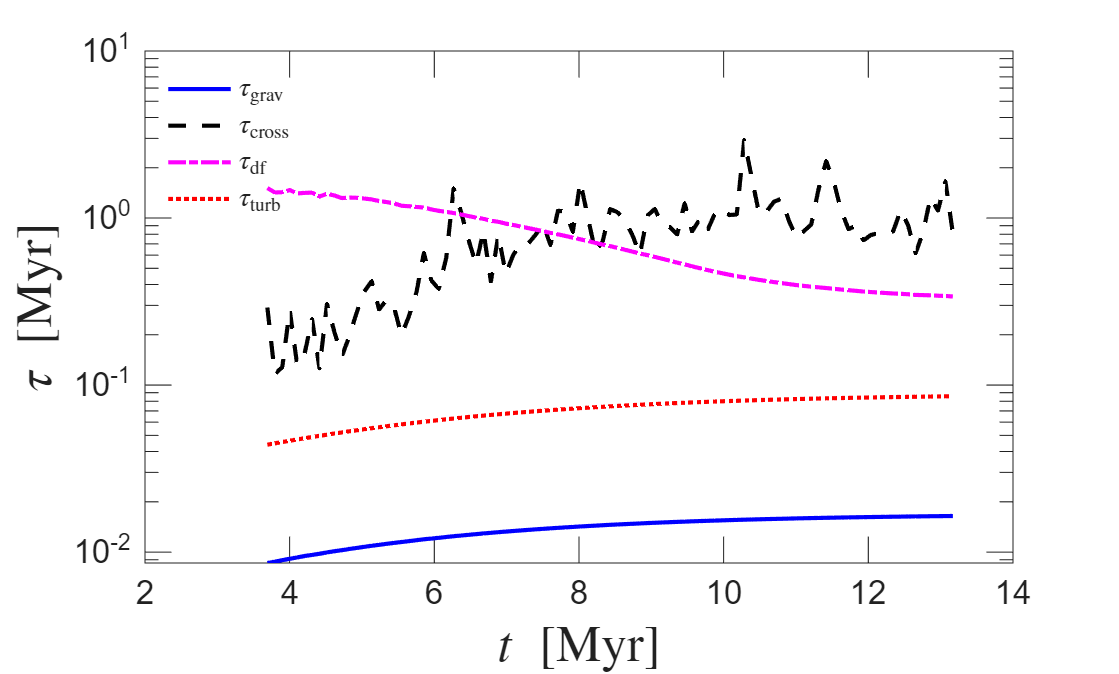}
			\caption{$m_{\bullet,{\rm f}} = 985\,{\rm M}_{\odot}$}
			\label{fig:stoch_timescales_t_m985}
		\end{subfigure}
		\caption{The evolution of timescales used in Eq. (\ref{eq:Deltat_stoch}) for two heavy BHs grown to final masses (a) $100\,{\rm M}_\odot$ and (b) $985\,{\rm M}_\odot$, in an indicative run of our typical cluster $M_{\star} = 10^6 \,{\rm M}_\odot$, $r_{c,\star} = 1 \,{\rm pc}$, $\varepsilon = 0.35$.}
		\label{fig:stoch_timescales_t_m}
	\end{figure}

	There is a hierarchy behind Eq. (\ref{eq:Deltat_stoch}). Dynamical friction is a cumulative relaxation process for each BH and therefore should operate on a timescale longer than the local dynamical time. 
	It is natural then to update the stochastic forcing on the timescale over which the local perturber population evolves, $\Delta t_{\rm stoch} = \mathscr{O}(\tau_{\rm grav})$, as long as $\tau_{\rm df} \gg\tau_{\rm grav}$. 
	In all our runs we have typically $\tau_{\rm grav} \lesssim 10^{-2} \tau_{\rm df}$, and at least $\tau_{\rm grav} < 10^{-1} \tau_{\rm df}$, throughout the whole time evolution, as illustrated in Fig. \ref{fig:stoch_timescales_t_m} for two represantive BHs of different mass.
	Accordingly, $\tau_{\rm grav}$ sets the stochastic timestep for $\sim 99\%$ of all kicks for BHs with final mass $m_{\bullet} \geq 15 {\rm M_{\odot}}$ and for the vast majority $\gtrsim 75\%$ of BHs with lower final masses. 
	The crossing time $\tau_{\rm cross}$ becomes relevant only for low-mass BHs displaced to the outskirts of the core after a previous strong kick, but still getting $\tau_{\rm cross} \approx \tau_{\rm grav}$ in those cases. The $\tau_{\rm turb}$ remains always larger than $\tau_{\rm grav}$ for our adopted turbulence Mach number $\mathcal{M}_{\rm turb} = 2$, and more generally for any $\mathcal{M}_{\rm turb} \lesssim 10$.                 
	We also implemented in the code prescribed bounds on $\Delta t_{\rm stoch}$ to improve numerical stability, which nevertheless were never violated by Eq. (\ref{eq:Deltat_stoch}) in practice.
	
	\begin{figure}[h!]
		\centering
		\includegraphics[width=0.9\columnwidth]{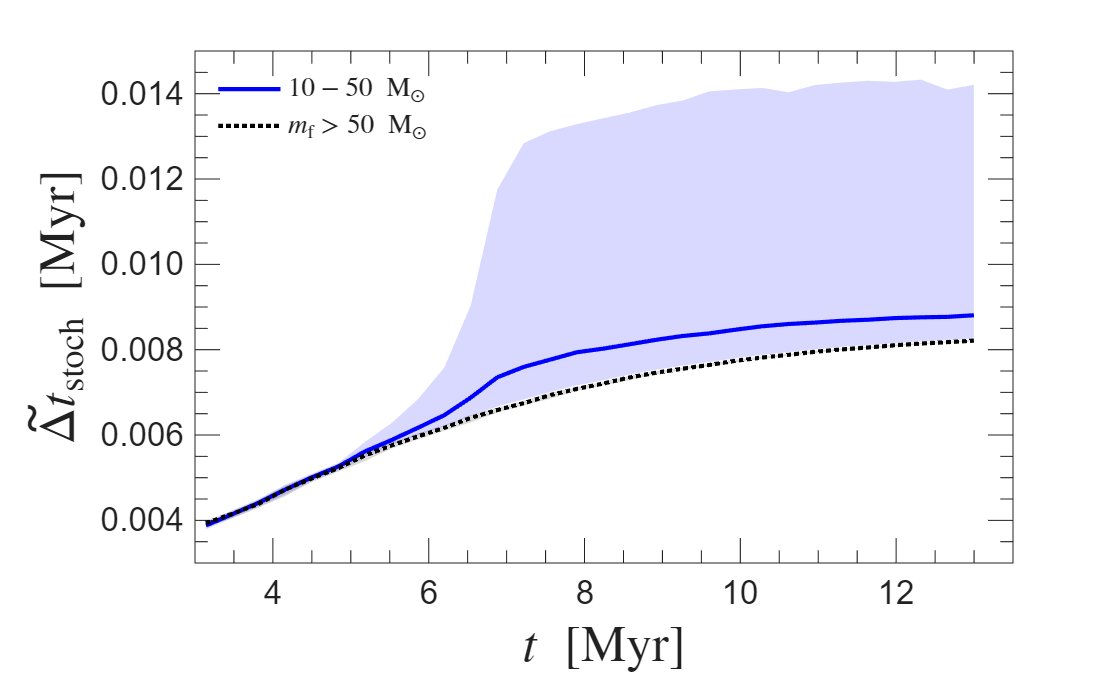}
		\caption{Evolution of the median stochastic timestep $\tilde{\Delta t}_{\rm stoch}$ and its $1\sigma$ percentile for final low-mass ($10-50\,{\rm M}_\odot$) and heavy ($> 50\,{\rm M}_\odot$) BHs. Two runs of a cluster as in Fig. \ref{fig:stoch_timescales_t_m}.}
		\label{fig:Dt_stoch_t_masses}
	\end{figure}
	
	Furthermore, in order to account for the approximate nature of $\Delta t_{\rm stoch}$ we introduce the numerical control parameter $f_{\rm dt}$. 
	For the simulations presented here, we adopt the value $f_{\rm dt} = 0.5$, resolving the stochastic updates more finely than the local dynamical timescale without oversampling the coarse-grained evolution. 
	We also explored the parameter range $f_{\rm dt} = 0.1 - 2.0$ and found that our primary results on the BH spin-mass correlation are robust under this variation. 
	In particular, the general spin-mass trend remains the same. The difference is that a longer timestep tends to broaden the spread of spin values resulting in a median saturation spin at high masses smaller by $\sim 10\%$ or less for $f_{\rm dt} = 2.0$ with respect to $f_{\rm dt} = 0.5$. Nevertheless, similar maximum spin values are achieved.
	
	Among heavy BHs with $m_{\bullet,{\rm f}} \geq 50\,{\rm M}_\odot$, the $\Delta t_{\rm stoch}$ presents negligible variation throughout the cluster evolution. Lower-mass BHs show a larger spread, because they receive larger velocity kicks and also are born at different times (originating in diverse stelar evolution channels). This is manifested as a sharp broadening of their distribution depicted in Fig. {\ref{fig:Dt_stoch_t_masses}}, where we plot the evolution of the median $\tilde{\Delta t}_{\rm stoch}$ and the $1\sigma$ percentile for light and heavy BHs. Still, for all BHs, it is $\Delta t_{\rm stoch} = \mathscr{O}(10^{-2}\,{\rm Myr})$ varying less than half an order of magnitude. This is true also for all cluster parameters inspected here, with lower compactness at the same cluster mass leading to slightly higher $\Delta t_{\rm stoch}$, but still of the same order of magnitude. 
	
	Finally, we remark that heavy BHs tend to receive more kicks than lower-mass BHs, having shorter stochastic timesteps and being born mostly earlier in cluster's life. For our typical massive compact cluster, $M_{\star} = 10^6\,{\rm M}_\odot$, $r_{c,\star} = 1\,{\rm pc}$, we report the median, the range and $1\sigma$ percentile (in parenthesis) of stochastic cycles for several masses as follows: (i) the BHs which have grown massive, $(150-10^3)\,{\rm M}_\odot$, have received $\approx 1400^{+180(130)}_{-200(120)}$ kicks,  
	(ii) the BHs with final mass within the mass gap, $(50-150)\,{\rm M}_\odot$, have received $\approx 1300^{+290(170)}_{-190(120)}$ kicks, 
	(iii) the low-mass BHs, $m_{\bullet,{\rm f}}< 50\,{\rm M}_\odot$, have received $\approx 600^{+990(300)}_{-500(260)}$ kicks. The primary cluster parameter that determines the stochastic cycles is the cluster compactness rather than the cluster mass. Less compact clusters that can drive BH mass growth only up to $m_{\bullet,{\rm max}} \lesssim 150\,{\rm M}_\odot$ generate $\sim 25\%$ less number of kicks per BH. 
	
	\section{Accretion disk timescales} \label{app:disk}
	
	\paragraph{Warp timescale.}
	The outer radius (\ref{eq:R_out}) of the accretion disk at which the warp propagates is defined by the condition
	$t_{\rm w}(R_{\rm out}) = \Delta t_{\rm stoch}$, where $t_{\rm w}(R)$ is the local warp-propagation timescale.
	For thin disks, warp communication occurs diffusively \citep{2002apa..book.....F}, 
	with a radial diffusion timescale \citep{1992MNRAS.258..811P} (see also \citealt{2007MNRAS.381.1287L})
	\begin{equation}\label{eq:t_w_thin}
		t_{\rm w}^{({\rm thin})}(R) \sim \frac{R^2}{\nu_2}
		= 2 {\rm s} \;
		\left(\tfrac{f_{\nu_2}}{0.6} \right)^{-1}
		\left(\tfrac{\alpha}{0.1} \right)^{\frac{6}{5}}
		\dot{m}^{-\frac{3}{10}} 
		\left(\tfrac{m_{\rm BH}}{50{\rm M}_\odot} \right)^{\frac{6}{5}}
		\left(\tfrac{R}{r_{\rm g}} \right)^{\frac{5}{4}}
	\end{equation}
	where $\nu_2$ is the transverse viscosity and $f_{\nu_2}$ a coefficient determined in simulations depending on $\alpha$ \citep{2009MNRAS.399.2249P}. We use the value $f_{\nu_2} = 0.6$ appropriate for the value $\alpha = 0.1$ that we use for thin disks. 
	
	For slim disks, which belong to the bending-wave regime ($\alpha < H/R$), warping disturbances propagate on a timescale \citep{1999MNRAS.309..929N}
	\begin{equation}\label{eq:t_w_slim}
		t_{\rm w}^{({\rm slim})}(R) \sim \frac{2R}{c_{s,{\rm disk}}}
		= 4.9\cdot 10^{-3} {\rm s} 
		\left( \tfrac{H/R}{0.1}\right)^{-1}
		\left( \tfrac{m_{\rm BH}}{50{\rm M}_\odot}\right)
		\left( \tfrac{R}{r_{\rm g}}\right)^{\frac{3}{2}}
	\end{equation}
	where $c_{s,{\rm disk}} \approx H \Omega_{\rm K,disk}(R)$.
	
	\paragraph{BH-alignment timescale.}
	We calculate the BH-alignment timescale ($t_{{\rm al},\bullet}$) that is the typical time required for the BH to align with the disk (case $J_{\rm d}/(2J_{\rm BH}) > 1$). This is determined by the gravitomagnetic timescale, $t_{\rm gm}$, of the torque exerted by the disk onto the BH
	\begin{equation}
		t_{\rm gm}(R) \sim \frac{c^2}{4\pi G } \frac{1}{\Sigma (R) \Omega_{{\rm K},\bullet}(R) R}
	\end{equation}
	at the distance of maximum torque $R_{\rm T}$. It is \citep{2007MNRAS.381.1617M}
	\begin{equation}\label{eq:t_al_bas}
		t_{{\rm al},\bullet} \sim \sqrt{2} t_{\rm gm}(R_{\rm T})
		=
		17.9{\rm yr} 
		\left(\tfrac{\alpha}{0.01} \right)^{\frac{4}{5}}
		\left(\tfrac{m_{\rm BH}}{50{\rm M}_\odot} \right)^{\frac{4}{5}}
		\left(\tfrac{R_{\rm T}}{r_{\rm g}} \right)^{\frac{5}{4}}
		\dot{m}^{-\frac{7}{10}} 
		.
	\end{equation}
	
	In the thin-disk case, the radius of maximum torque is the ``warp'' or ``Bardeen--Petterson'' radius defined as the distance at which the warp timescale (\ref{eq:t_w_thin}) equals the Lense--Thirring precession timescale $\Omega_{\rm LT}^{-1}$, that is $R_{\rm T}^{({\rm thin})} \sim \frac{4\pi G J_{\rm BH}}{\nu_2 c^2}$ which gives
	\begin{equation}\label{eq:R_T_thin}
		\frac{R_{\rm T}^{({\rm thin})} } {r_{\rm g}} \sim  374.3 \, a_{*}^{\frac{4}{7}}
		\left(\tfrac{f_{\nu_2}}{0.6} \right)^{-\frac{4}{7}}
		\left(\tfrac{\alpha}{0.1} \right)^{\frac{24}{35}}
		\dot{m}^{-\frac{6}{35}} 
		\left(\tfrac{m_{\rm BH}}{50{\rm M}_\odot} \right)^{\frac{4}{35}}.
	\end{equation} 
	Substituting in (\ref{eq:t_al_bas}) we get
	\begin{equation}\label{eq:t_al_thin}
		\frac{t_{{\rm al},\bullet}^{({\rm thin)}}}{\Delta t_{\rm stoch}} \sim 18 \, a_{*}^{\frac{5}{7}}
		\left(\tfrac{\Delta t_{\rm stoch}}{10^{-2}{\rm Myr}} \right)^{-1}                    
		\left(\tfrac{f_{\nu_2}}{0.6} \right)^{-\frac{5}{7}}
		\left(\tfrac{\alpha}{0.1} \right)^{\frac{58}{35}}
		\left(\tfrac{m_{\rm BH}}{50{\rm M}_\odot} \right)^{\frac{-2}{35}}
		\dot{m}^{-\frac{32}{35}} . 
	\end{equation}
	In practice, this timescale is mostly irrelevant since the case $J_{\rm d}/(2J_{\rm BH}) > 1$ refers primarily to more massive BHs for which $\dot{m}> 0.3$, i.e. slim disks (see Eq. \ref{eq:J_ratio}).
	
	In the slim-disk case (bending-wave $\alpha < H/R$) the maximum torque is applied at a distance \citep{2005MNRAS.363...49K}
	\begin{equation}
		\frac{R_{\rm T}^{({\rm slim})}}{r_{\rm g}} \sim 3.6\, a_{*}^{\frac{2}{5}} (H/R)^{-\frac{4}{5}} .
	\end{equation} 
	Substituting in (\ref{eq:t_al_bas}) we get
	\begin{equation}\label{eq:t_al_slim}
		\frac{t_{{\rm al},\bullet}^{({\rm slim})}}{\Delta t_{\rm stoch}} \sim
		0.09 a_{*}^{\frac{1}{2}}
		\left(\tfrac{\Delta t_{\rm stoch}}{10^{-2}{\rm Myr}} \right)^{-1}                    
		\left(\tfrac{H/R}{0.1} \right)^{-1}
		\left(\tfrac{\alpha}{0.01} \right)^{\frac{4}{5}}
		\left(\tfrac{m_{\rm BH}}{50{\rm M}_\odot} \right)^{\frac{4}{5}}
		\dot{m}^{-\frac{7}{10}} .
	\end{equation}
	We see that massive BHs within the mass gap, $m_{\rm BH} \lesssim 150 {\rm M}_\odot$, 
	satisfy $t_{{\rm al},\bullet}^{({\rm slim})} \lesssim 0.1 \Delta t_{\rm stoch}$ 
	even at high spin $a_{*} \approx 0.9$ and sufficiently high accretion rate $\dot{m} \gtrsim 1$.

	Note that when the BH dominates the angular momentum, $J_{\rm d}/(2J_{\rm BH}) < 1$, it is the inner disk which reorients itself to (or counter to) the BH spin. In this case, Eqs. (\ref{eq:t_w_thin}) or (\ref{eq:t_w_slim}), for the thin or slim disk respectively, should be used (instead of Eq. \ref{eq:t_al_bas}) substituting $R$ with an inner radius $R_{\rm in} \approx R_{\rm ISCO}$. This yields $t_{\rm al,d} \ll \Delta t_{\rm stoch}$ for the parameter space of our study.
	
\end{document}